\documentclass[]{emulateapj}

\begin{document}

\newcommand{\kms}{km s$^{-1}$}
\newcommand{\msun}{$M_{\odot}$}
\newcommand{\rsun}{$R_{\odot}$}
\newcommand{\teff}{$T_{\rm eff}$}
\newcommand{\logg}{$\log{g}$}

\title{The ELM Survey. V. Merging Massive White Dwarf Binaries\altaffilmark{*}}

\author{Warren R.\ Brown$^1$,
        Mukremin Kilic$^2$,
        Carlos Allende Prieto$^{3,4}$,
        A.\ Gianninas$^2$
        and Scott J.\ Kenyon$^1$
        }

\affil{ $^1$Smithsonian Astrophysical Observatory, 60 Garden St, Cambridge, MA 02138 USA\\
        $^2$Homer L. Dodge Department of Physics and Astronomy, University of Oklahoma, 440 W. Brooks St., Norman, OK, 73019 USA\\
        $^3$Instituto de Astrof\'{\i}sica de Canarias, E-38205, La Laguna, Tenerife, Spain\\
        $^4$Departamento de Astrof\'{\i}sica, Universidad de La Laguna, E-38206 La Laguna, Tenerife, Spain
        }

\email{wbrown@cfa.harvard.edu, kilic@ou.edu}

\altaffiltext{*}{Based on observations obtained at the MMT Observatory, a joint
facility of the Smithsonian Institution and the University of Arizona.}

\shorttitle{ Extremely Low Mass White Dwarf Survey. V. }
\shortauthors{Brown et al.}

\begin{abstract}

        We present the discovery of 17 low mass white dwarfs (WDs) in short-period
$P\le1$ day binaries.  Our sample includes four objects with remarkable
$\log{g}\simeq5$ surface gravities and orbital solutions that require them to be
double degenerate binaries.  All of the lowest surface gravity WDs have metal lines
in their spectra implying long gravitational settling times or on-going accretion.  
Notably, six of the WDs in our sample have binary merger times $<$10 Gyr.  Four have
$\gtrsim$0.9 \msun\ companions.  If the companions are massive WDs, these four
binaries will evolve into stable mass transfer AM CVn systems and possibly explode
as underluminous supernovae.  If the companions are neutron stars, then these may be
milli-second pulsar binaries.  These discoveries increase the number of detached,
double degenerate binaries in the ELM Survey to 54; 31 of these binaries will merge
within a Hubble time.

\end{abstract}


\section{INTRODUCTION}

        Extremely low mass (ELM) WDs, degenerate objects with $\log{g}<7$ (cm
s$^{-2}$) surface gravity or $\lesssim$0.3 \msun\ mass, are the product of common
envelope binary evolution \citep[e.g.][]{marsh95}.  ELM WDs are thus the signposts
of the type of binaries that are strong gravitational wave sources and possible
supernovae progenitors.  The goal of the ELM Survey is to discover and characterize
the population of ELM WDs in the Milky Way.

	Previous ELM Survey papers have reported the discovery of 40 WDs spanning
0.16 \msun\ to 0.49 \msun\ found in the Hypervelocity Star (HVS) Survey, the Sloan
Digital Sky Survey (SDSS), and in our own targeted survey \citep{brown10c, brown11a,
brown12a, kilic10, kilic11a, kilic12a}.  We refer to this full sample of WDs as the
ELM Survey sample, but reserve the term ``ELM WD'' for those objects with
$\log{g}<7$.  All of our WDs are found in short-period, detached binaries, 60\% of
which have merger times $<$10 Gyr.  Three notable systems are detached binaries with
$<$40 min orbital periods \citep{kilic11c, kilic11b}.  The eclipsing system
J0651+2844 is the second-strongest gravitational wave source in the mHz range
\citep{brown11b}. We measured its period change in one year with optical eclipse
timing \citep{hermes12c}.  Other results from the ELM Survey include the first
tidally distorted WDs \citep{kilic11b, hermes12a} and the first pulsating
helium-core WDs \citep{hermes12b, hermes12d}.

        Here we present the discovery of 17 new WD binaries identified from spectra 
previously obtained for the HVS Survey of \citet{brown05, brown06, brown06b, 
brown07a, brown07b, brown09a, brown12b}.  \citet{kilic07} analyzed the 
visually-identified WDs in the original dataset and discovered one ELM WD binary 
\citep{kilic07b}. This approach failed to identify the lowest surface gravity WDs.  
We now fit stellar atmosphere models to the entire collection of spectra not 
previously identified as WDs, and acquire follow-up spectroscopy of new ELM WD 
candidates.  The result of this effort is that we find low surface gravity objects 
that might not be considered WDs if not for their observed orbital motion.

	We chose to call objects with $5<\log{g}<7$ ``ELM WDs'' because these
objects occupy a unique region of surface gravity/effective temperature space that
overlaps the terminal WD cooling branch for ELM WDs (see Figure \ref{fig:teff}).
This region is well separated from both hydrogen-burning main sequence tracks and
helium-burning horizontal branch tracks.  Our objects are systematically
$\sim$10,000 K too cool, given their surface gravities, to be helium burning sdB
stars.  \citet{kaplan13} refer to a similar type of object as a ``proto-WD.'' The
lowest gravity objects in our sample may indeed have hydrogen shell burning and thus
are, properly speaking, proto-WDs, but here we address our sample of low gravity
objects as ELM WDs.  A variety of observations demonstrate that degenerate
helium-core WDs exist at our observed temperatures and surface gravities.  ELM WD
companions to milli-second pulsars are directly observed \citep[e.g.][]{bassa06,
cocozza06} at the temperatures and gravities targeted by the ELM Survey.  The
measured radius of the $\log{g}=6.67\pm0.04$ object in J0651+2844, a
$0.0371\pm0.0012$ \rsun\ star, demonstrates it is a degenerate WD \citep{brown11b,
hermes12c}.  \citet{vangrootel13} account for the unusually long pulsation periods
of the ELM WD pulsators with low mass WD models.  Hence, calling our low surface
gravity objects ELM WDs is appropriate.

	Interestingly, we find ELM WDs in binaries with $>$0.9 \msun\ companions and
rapid merger times.  When these detached binaries begin mass transfer,
\citet{marsh04} show that the extreme mass ratios will lead to stable mass transfer.  
For the case of massive WD accretors, theorists predict large helium flashes that
may ignite thermonuclear transients dubbed ``.Ia'' supernovae \citep{bildsten07,
shen09} or may detonate the surface helium-layer and the massive WD \citep{nomoto82,
woosley94, sim12} and produce an underluminous supernova. However, the final outcome
of such mergers is uncertain and they may not trigger supernovae explosions
\citep{dan12}.  If the companions are instead neutron stars, this would be the first
time a milli-second pulsar is identified through its low-mass WD companion. Such
systems allow measurement of the binary mass ratio and the neutron star mass through
a combination of the pulsar orbit obtained from radio timing and the WD orbit
obtained from optical radial velocity observations.

        The importance of identifying the ELM WDs in the HVS Survey is that the 
HVS Survey is a well-defined and a nearly 100\% complete spectroscopic survey. 
With a complete sample of ELM WDs we can measure the space density, period 
distribution, and merger rate of ELM WDs, and link ELM WD merger products to 
populations of AM CVn stars, R CrB stars, and possibly underluminous 
supernovae.  In a stellar evolution context, our ELM WD survey complements studies 
of WD binaries with main sequence companions \citep[e.g.][]{pyrzas12, rebassa12} 
and with sdB star companions \citep[e.g.][]{geier12, silvotti12}.  sdB 
stars are \teff $>25,000$ K helium-burning precursors to WDs \citep{heber09}.
Our ELM WDs, on the other hand, have \teff $<20,000$ K.

        We organize this paper as follows.  In Section 2 we discuss our observations 
and data analysis.  In Section 3 we present the orbital solutions for 17 new ELM WD 
binaries.  In Section 4 we discuss the properties of the ELM WD sample.  We conclude 
in Section 5.

\section{DATA AND ANALYSIS}

\subsection{Target Selection}

	The HVS Survey is a targeted spectroscopic survey of $15<g_0<20$ stars with 
the colors of $\simeq$3 \msun\ main sequence stars, stars which should not exist at 
faint magnitudes in the halo unless they were ejected there.  The target selection 
is detailed in \citet{brown12b} and spans $-0.4 < (g-r)_0 \lesssim -0.25$, $0.4 
\lesssim (u-g)_0 < 1.07$.  This color selection fortuitously targets WDs in the 
approximate range $10,000<$ \teff\ $<20,000$ K and $\log{g} \lesssim 7.5$.

        We separate ELM WDs from the other stars in the HVS Survey using stellar
atmosphere model fits to the single-epoch spectra.  Our initial set of fits uses an
upgraded version of the {\sc ferre} code described by \citet{allende06} and
synthetic DA WD pure hydrogen spectra kindly provided by D.\ Koester.  The grid of
WD model atmospheres covers effective temperatures from 6000 K to 30,000 K in steps
of 500 K to 2000 K, and surface gravities from \logg=5.0 to 9.0 in steps of 0.25
dex.  The model atmospheres are calculated assuming local thermodynamic equilibrium
and include both convective and radiative transport \citep{koester08}.

        We fit 2,000 spectra, mostly from the original HVS Survey \citep{brown05,
brown06, brown06b, brown07a, brown07b, brown09a} but also some newly acquired data
\citep{brown12b}.  These 2,000 spectra were not previously analyzed for the ELM
Survey because the spectra were not previously identified as WDs.  We fit both
flux-calibrated spectra (for improved \teff\ constraints) as well as
continuum-corrected Balmer line profiles (insensitive to reddening and flux
calibration errors).  For the continuum-corrected Balmer line profiles, we normalize
the spectra by fitting a low-order polynomial to the regions between the Balmer
lines.  We adopt the parameters from the flux-calibrated spectra, except in cases
where the spectra were obtained in non-photometric conditions.  The uncertainties in
our single-epoch measurements are typically $\pm$500 K in \teff\ and $\pm0.1$ dex in
\logg.  From these fits we identify 57 low mass WD candidates.

\begin{figure}          
 \plotone{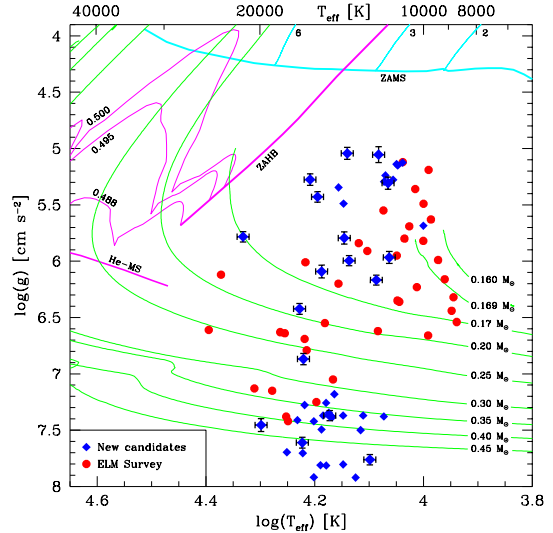}
 \caption{ \label{fig:teff}
        Surface gravity vs.\ effective temperature of the 57 WD candidates (blue 
diamonds) and the previously published ELM Survey WDs (red circles).  The 17 newly 
discovered binaries are plotted with errorbars and parameters obtained from 
Gianninas stellar atmosphere models; the parameters of the other WDs were obtained 
from Koester models.  For reference, we plot theoretical tracks for 0.16--0.45 
\msun\ hydrogen atmosphere WDs \citep[green lines,][]{panei07}, main sequence tracks 
for 2, 3 and 6 \msun\ stars \citep[cyan lines,][]{girardi04}, and horizontal branch 
tracks for 0.488, 0.495, and 0.500 \msun\ stars \citep[magenta lines,][]{dorman93}.  
Zero-age main sequence (ZAMS) and zero-age horizontal branch (ZAHB) isochrones are 
drawn with thick lines, as is the homogenous helium-burning main sequence (He-MS) 
from \citet{paczynski71}.}
 \end{figure}

\subsection{New Spectroscopic Observations}

        We obtain follow-up spectra for each of the candidate low mass WDs to
improve stellar atmosphere parameters and to search for velocity variability.

        Observations were obtained over the course of seven observing runs at the
6.5m MMT telescope between March 2011 and February 2013.  We used the Blue Channel
spectrograph \citep{schmidt89} with the 832 line mm$^{-1}$ grating, which provides a
wavelength coverage 3650 \AA\ to 4500 \AA\ and a spectral resolution of 1.0 \AA.  
All observations were paired with a comparison lamp exposure, and were
flux-calibrated using blue spectrophotometric standards \citep{massey88}. The
extracted spectra typically have a signal-to-noise (S/N) of 7 per pixel in the
continuum and a 14 \kms\ radial velocity error.

        We obtained additional spectroscopy for $g<17$ mag ELM WD candidates in
queue scheduled time at the 1.5m FLWO telescope. We used the FAST spectrograph
\citep{fabricant98} with the 600 line mm$^{-1}$ grating and a 1.5\arcsec\ slit,
providing wavelength coverage 3500 \AA\ to 5500 \AA\ and a spectral resolution of
1.7 \AA.  All observations were paired with a comparison lamp exposure, and were
flux-calibrated using blue spectrophotometric standards.  The extracted spectra
typically have a S/N of 10 per pixel in the continuum and a 18 \kms\ radial velocity
error.
	
\subsection{ELM WD Identifications}

	Our follow-up observations provide improved stellar atmosphere constraints, 
from which we determine that 24 (42\%) of the candidates are probable ELM WDs with 
$5<\log{g}<7$ (see Figure \ref{fig:teff}).  The remaining candidates have either 
$\log{g}>7$ and are normal DA WDs, or $\log{g}\le5$ and are presumably halo blue 
horizontal branch or blue straggler stars.

        Figure \ref{fig:teff} plots the distribution of \teff\ vs.\ \logg\ for all
of the WDs with $\log{g}>5$.  Previously published ELM Survey stars are marked with
red circles.  The observed WDs overlap tracks based on \citet{panei07} models for
He-core WDs with hydrogen shell burning \citep[(green lines)][]{kilic10}, but do not
overlap main sequence nor horizontal branch evolutionary tracks. For reference, we
also plot \citet{girardi02, girardi04} solar metallicity main sequence tracks for 2,
3, and 6 \msun\ stars (cyan lines).  A helium-burning horizontal branch star can
have a surface gravity similar to a degenerate He WD \citep[e.g.,][]{heber03}, but
only at a systematically higher effective temperature than that targeted by the ELM
Survey.  This is illustrated by \citet{dorman93} [Fe/H]=-1.48 horizontal branch
tracks for 0.488, 0.495, and 0.500 \msun\ stars (magenta lines), as well as the
homogenous helium-burning main sequence from \citet{paczynski71}.  The zero-age main
sequence and helium-burning horizontal branch isochrones (thick lines) mark the
limits.

	Seventeen of the WDs show significant velocity variability, including twelve
of the newly identified ELM WDs.  The other ELM WDs have insufficient coverage for
detecting (or ruling out) velocity variability.  We focus the remainder of this
paper on the 17 well-constrained systems.

\begin{figure}          
 \plottwo{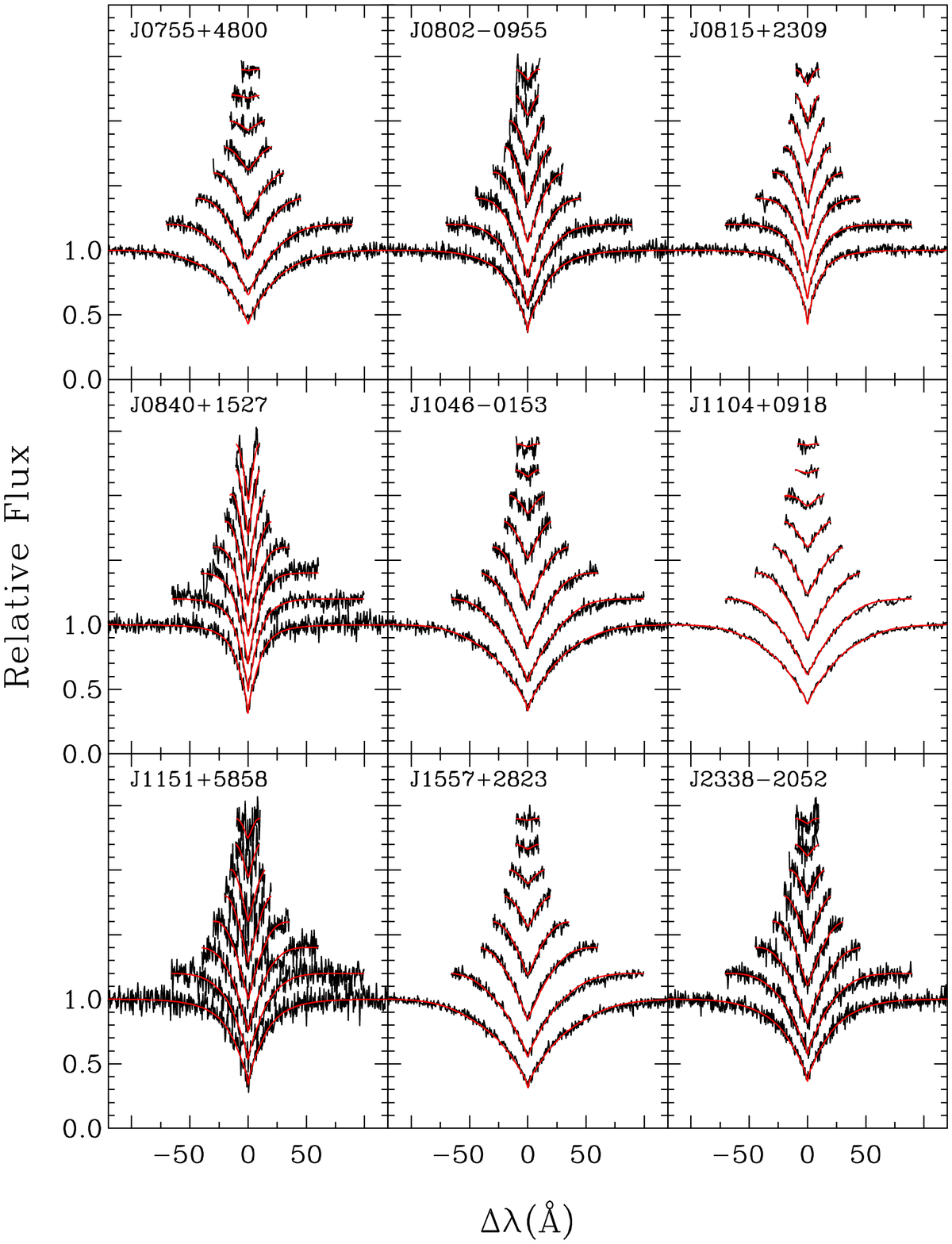}{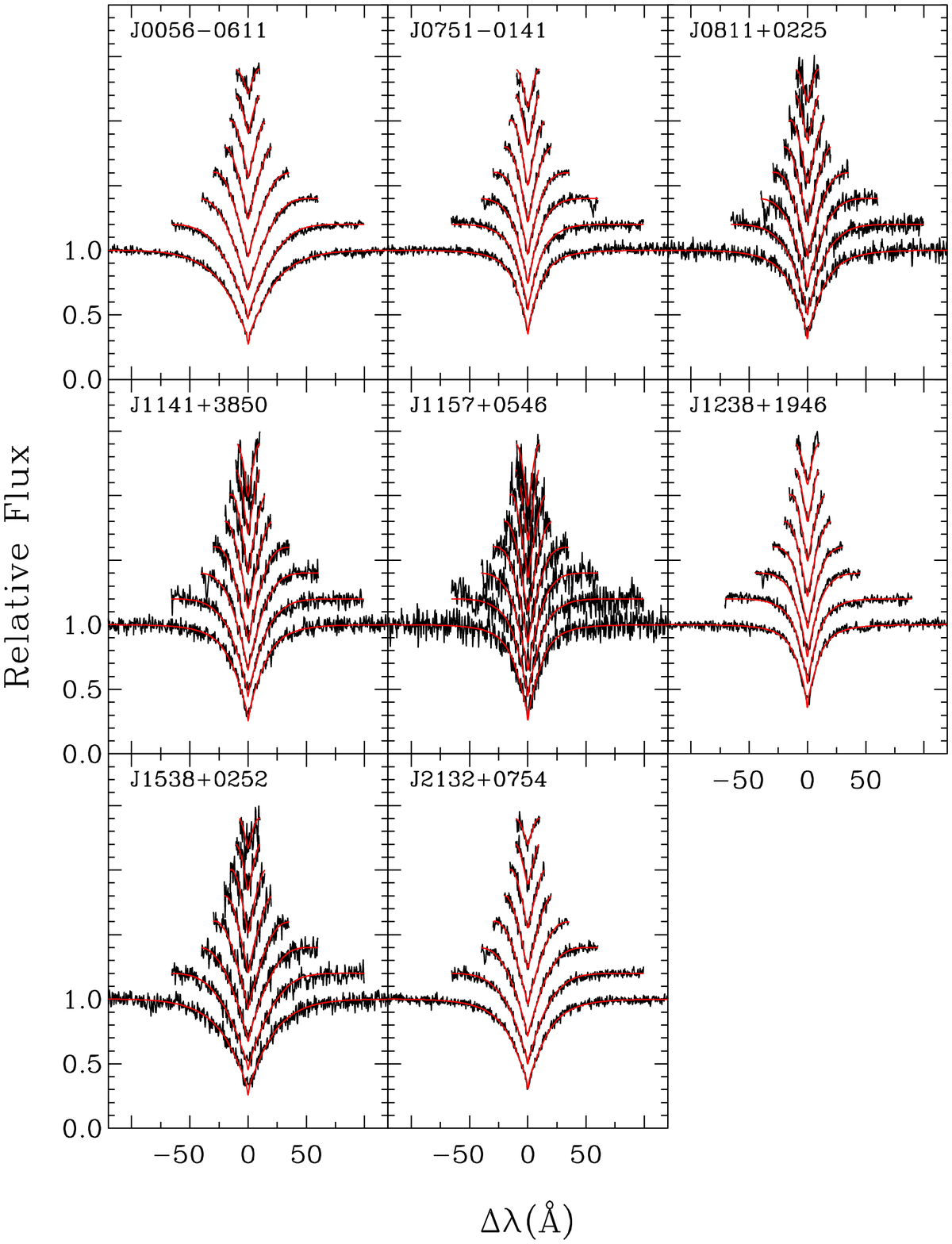}
 \caption{ \label{fig:spec}
	Gianninas model fits ({smooth red lines}) overplotted on the composite
observed spectra ({black lines}) for the 17 WD binaries.  The Balmer lines are
arranged from H12 (top) to H$\gamma$ (bottom); left panel plots the $K<200$ \kms\
binaries, right panel plots the $K>200$ \kms\ binaries. }
 \end{figure}

\subsection{Improved Atmosphere Parameters}

        Our sample of 17 well-constrained systems contains four \logg\ $\simeq5$
objects, objects at the edge of our stellar atmosphere grid that might not be
considered degenerate WDs if not for their observed orbital motion.  We also fit
these stars to grids of \citet{kurucz93} models spanning $-4 \le {\rm [Fe/H]} \le
0.5$, $1 \le \log{g} \le 5$ \citep[see][]{allende08} but found very poor solutions.  
This motivated us to perform a new set of stellar atmosphere fits using
\citet{gianninas11} hydrogen model atmospheres.  The Gianninas models employ
improved Stark broadening profiles \citep{tremblay09}, and the ML2/$\alpha$ = 0.8
prescription of the mixing-length theory for models where convective energy
transport is important \citep[][]{tremblay10}.  We calculate atmosphere models for
low surface gravities, and use a grid that covers \teff\ from 4000 K to 30,000 K in
steps ranging from 250 to 5000 K and \logg\ from 5.0 to 8.0 in steps of 0.25 dex.

\begin{deluxetable*}{ccccccccc}    
\tabletypesize{\scriptsize}
\tablecolumns{9}
\tablewidth{0pt}
\tablecaption{White Dwarf Physical Parameters\label{tab:param}}
\tablehead{
        \colhead{Object} & 
	\colhead{RA} & 
	\colhead{Dec} &
        \colhead{$T_{\rm eff}$} &
        \colhead{$\log g$} &
        \colhead{Mass} &
        \colhead{$g_0$} &
        \colhead{$M_g$} &
        \colhead{$d_{helio}$} \\
   & (h:m:s) & (d:m:s) & (K) & (cm s$^{-2}$) & (\msun) & (mag) & (mag) & (kpc)
}
        \startdata
J0056$-$0611 &  0:56:48.232 &  -6:11:41.62 & $12210 \pm 180$ & $6.167 \pm 0.044$ & 0.17 & $17.208 \pm 0.023$ &  8.0 & 0.69 \\
J0751$-$0141 &  7:51:41.179 &  -1:41:20.90 & $15660 \pm 240$ & $5.429 \pm 0.046$ & 0.17 & $17.376 \pm 0.015$ &  8.0 & 0.75 \\
J0755$+$4800 &  7:55:19.483 &  48:00:34.07 & $19890 \pm 350$ & $7.455 \pm 0.057$ & 0.42 & $15.878 \pm 0.019$ &  9.7 & 0.18 \\
J0802$-$0955 &  8:02:50.134 &  -9:55:49.84 & $16910 \pm 280$ & $6.423 \pm 0.048$ & 0.20 & $18.604 \pm 0.012$ &  8.2 & 1.19 \\
J0811$+$0225 &  8:11:33.560 &   2:25:56.76 & $13990 \pm 230$ & $5.794 \pm 0.054$ & 0.17 & $18.569 \pm 0.013$ &  8.0 & 1.30 \\
J0815$+$2309 &  8:15:44.242 &  23:09:04.92 & $21470 \pm 340$ & $5.783 \pm 0.046$ & 0.17 & $17.623 \pm 0.015$ &  6.7 & 1.53 \\
J0840$+$1527 &  8:40:37.574 &  15:27:04.53 & $13810 \pm 240$ & $5.043 \pm 0.053$ & 0.17 & $19.141 \pm 0.018$ &  8.0 & 1.69 \\
J1046$-$0153 & 10:46:07.875 &  -1:53:58.48 & $14880 \pm 230$ & $7.370 \pm 0.045$ & 0.37 & $17.927 \pm 0.020$ & 10.2 & 0.36 \\
J1104$+$0918 & 11:04:36.739 &   9:18:22.74 & $16710 \pm 250$ & $7.611 \pm 0.049$ & 0.46 & $16.543 \pm 0.016$ & 10.3 & 0.18 \\
J1141$+$3850 & 11:41:55.560 &  38:50:03.02 & $11620 \pm 200$ & $5.307 \pm 0.054$ & 0.17 & $18.972 \pm 0.018$ &  8.0 & 1.56 \\
J1151$+$5858 & 11:51:38.381 &  58:58:53.22 & $15400 \pm 300$ & $6.092 \pm 0.057$ & 0.17 & $20.046 \pm 0.033$ &  8.0 & 2.57 \\
J1157$+$0546 & 11:57:34.455 &   5:46:45.58 & $12100 \pm 250$ & $5.054 \pm 0.071$ & 0.17 & $19.798 \pm 0.021$ &  8.0 & 2.29 \\
J1238$+$1946 & 12:38:00.096 &  19:46:31.45 & $16170 \pm 260$ & $5.275 \pm 0.051$ & 0.17 & $17.155 \pm 0.019$ &  8.0 & 0.68 \\
J1538$+$0252 & 15:38:44.220 &   2:52:09.49 & $11560 \pm 220$ & $5.967 \pm 0.053$ & 0.17 & $18.528 \pm 0.015$ &  8.0 & 1.28 \\
J1557$+$2823 & 15:57:08.483 &  28:23:36.02 & $12550 \pm 200$ & $7.762 \pm 0.046$ & 0.49 & $17.496 \pm 0.029$ & 11.2 & 0.18 \\
J2132$+$0754 & 21:32:28.360 &   7:54:28.24 & $13700 \pm 210$ & $5.995 \pm 0.045$ & 0.17 & $17.904 \pm 0.019$ &  8.0 & 0.96 \\
J2338$-$2052 & 23:38:21.505 & -20:52:22.76 & $16630 \pm 280$ & $6.869 \pm 0.050$ & 0.27 & $19.577 \pm 0.035$ &  9.0 & 1.29 \\
        \enddata
\end{deluxetable*}

\begin{deluxetable*}{lccrrcccc}
\tabletypesize{\scriptsize}
\tablecolumns{9}
\tablewidth{0pt}
\tablecaption{Binary Orbital Parameters\label{tab:orbit}}
\tablehead{
\colhead{Object}&
\colhead{$N_{obs}$}&
\colhead{$P$}&
\colhead{$K$}&
\colhead{$\gamma$}&
\colhead{Spec.\ Conjunction}&
\colhead{$M_2$}&
\colhead{$q$}&
\colhead{$\tau_{\rm merge}$}\\
 & & (days) & (km s$^{-1}$) & (km s$^{-1}$) & HJD-2450000 (days) & (\msun ) & & (Gyr)
}
        \startdata
J0056$-$0611 & 33 & $0.04338 \pm 0.00002$ & $376.9 \pm  2.4$ & $   4.2 \pm  1.8$ & $5864.76305 \pm 0.00008$ & $\ge$0.45 & $\le$0.374 & $\le$0.12 \\
J0751$-$0141 & 31 & $0.08001 \pm 0.00279$ & $432.6 \pm  2.3$ & $  61.9 \pm  1.8$ & $5623.60639 \pm 0.00013$ & $\ge$0.94 & $\le$0.182 & $\le$0.37 \\
J0755$+$4800 & 26 & $0.54627 \pm 0.00522$ & $194.5 \pm  5.5$ & $  42.6 \pm  3.8$ & $3730.89702 \pm 0.00208$ & $\ge$0.90 & $\le$0.468 & $\le$28 \\
J0802$-$0955 & 20 & $0.54687 \pm 0.00455$ & $176.5 \pm  4.5$ & $  27.0 \pm  3.4$ & $5623.34011 \pm 0.00228$ & $\ge$0.57 & $\le$0.348 & $\le$79 \\
J0811$+$0225 & 24 & $0.82194 \pm 0.00049$ & $220.7 \pm  2.5$ & $  77.4 \pm  1.9$ & $6329.61559 \pm 0.00296$ & $\ge$1.20 & $\le$0.142 & $\le$160 \\
J0815$+$2309 & 21 & $1.07357 \pm 0.00018$ & $131.7 \pm  2.6$ & $ -37.0 \pm  2.6$ & $5623.68831 \pm 0.00415$ & $\ge$0.47 & $\le$0.361 & $\le$620 \\
J0840$+$1527 & 19 & $0.52155 \pm 0.00474$ & $ 84.8 \pm  3.1$ & $  10.7 \pm  2.3$ & $4822.83144 \pm 0.00244$ & $\ge$0.15 & $\le$0.879 & $\le$230 \\
J1046$-$0153 & 16 & $0.39539 \pm 0.10836$ & $ 80.8 \pm  6.6$ & $ -33.3 \pm  4.6$ & $4597.54154 \pm 0.00293$ & $\ge$0.19 & $\le$0.509 & $\le$48 \\
J1104$+$0918 & 25 & $0.55319 \pm 0.00502$ & $142.1 \pm  6.0$ & $  72.5 \pm  4.0$ & $3793.84175 \pm 0.00269$ & $\ge$0.55 & $\le$0.831 & $\le$39 \\
J1141$+$3850 & 17 & $0.25958 \pm 0.00005$ & $265.8 \pm  3.5$ & $ -11.2 \pm  2.2$ & $3881.73870 \pm 0.00037$ & $\ge$0.76 & $\le$0.225 & $\le$10 \\
J1151$+$5858 & 17 & $0.66902 \pm 0.00070$ & $175.7 \pm  5.9$ & $  12.0 \pm  4.2$ & $5622.52148 \pm 0.00269$ & $\ge$0.61 & $\le$0.277 & $\le$150 \\
J1157$+$0546 &  9 & $0.56500 \pm 0.01925$ & $158.3 \pm  4.9$ & $-124.8 \pm  3.3$ & $4235.73743 \pm 0.00244$ & $\ge$0.44 & $\le$0.382 & $\le$120 \\
J1238$+$1946 & 21 & $0.22275 \pm 0.00009$ & $258.6 \pm  2.5$ & $  -6.6 \pm  1.2$ & $4236.71749 \pm 0.00032$ & $\ge$0.64 & $\le$0.266 & $\le$7.5 \\
J1538$+$0252 & 16 & $0.41915 \pm 0.00295$ & $227.6 \pm  4.9$ & $-157.9 \pm  3.9$ & $5385.57030 \pm 0.00146$ & $\ge$0.76 & $\le$0.222 & $\le$35 \\
J1557$+$2823 & 24 & $0.40741 \pm 0.00294$ & $131.2 \pm  4.2$ & $  10.4 \pm  3.0$ & $3563.72487 \pm 0.00127$ & $\ge$0.43 & $\le$0.886 & $\le$20 \\
J2132$+$0754 & 35 & $0.25056 \pm 0.00002$ & $297.3 \pm  3.0$ & $ -12.1 \pm  2.0$ & $5862.68632 \pm 0.00049$ & $\ge$0.95 & $\le$0.179 & $\le$7.7 \\
J2338$-$2052 & 25 & $0.07644 \pm 0.00712$ & $133.4 \pm  7.5$ & $   5.2 \pm  4.8$ & $5862.75190 \pm 0.00049$ & $\ge$0.15 & $\le$0.554 & $\le$0.95 \\
        \enddata 

\tablecomments{ 
Objects with significant period aliases: J0755+4800 (0.349 days),
J0840+1527 (0.340 days), J1046$-$0153 (0.659 days), J1104+0918 (0.355 days),
J1157+0546 (1.23 days), J1538+0252 (0.295 days), and J1557+2823 (0.677 and 0.290
days) as seen in Figure \ref{fig:pdm}. }

\end{deluxetable*}

        Our stellar atmosphere fits use the so-called spectroscopic technique 
described in \citet{gianninas11}.  One difference between our work and 
\citet{gianninas11} is that we fit higher-order Balmer lines, up to and including 
H12, observed in the low surface gravity ELM WDs.  The higher-order Balmer lines are 
sensitive to \logg\ and improve our surface gravity measurement.  For the handful of 
WDs in our sample with $\log{g}>7$, we only fit Balmer lines up to and including 
H10 since the higher order Balmer lines are not observed at higher surface gravity.

        The lowest gravity WDs in our sample show Ca and Mg lines in their spectra.  
For reference, the diffusion timescale for Ca in a $T_{\rm eff}=$ 10,000 K, 0.2
\msun\ H-rich WD is $\sim$$10^4$ yr \citep{paquette86}.  Extreme horizontal branch
stars, which can have surface gravities comparable to the lowest gravity WDs, have
longer $\sim$$10^6$ yr diffusion timescales \citep{michaud08}.  These diffusion
timescales are shorter than the WD evolutionary timescale, suggesting there may be
on-going accretion in these ultra-compact binary systems.  Near- and mid-infrared
observations are needed constrain the possibility of accretion.  We defer a detailed
analysis of the metal abundances in ELM WDs to a future paper. For our present
analysis we exclude the wavelength ranges where the metal lines are present in our
fits.

        Our error estimates combine the internal error of the model fits, obtained
from the covariance matrix of the fitting algorithm, and the external error,
obtained from multiple observations of the same object.  Uncertainties are typically
1.2\% in \teff\ and 0.038 dex in \logg\ \citep[see][for details]{liebert05}.  A
measure of the systematic uncertainty inherent in the stellar atmosphere models and
fitting routines comes from \citet{gianninas11}, who find a systematic uncertainty
of $\approx0.1$ dex in \logg.  This is corroborated by the difference we observe
between the different fitting methods discussed in Section 2.1 and here: the mean
difference and the dispersion in \teff is $1.1\% \pm 4.3\%$ and in \logg is $0.05
\pm 0.12$ dex. Figure \ref{fig:teff} plots the properties of the 17 WDs with their
internal errorbars.

	We use \citet{panei07} evolutionary tracks to estimate WD mass and
luminosity. For purpose of discussion, we assume $\log{g}<6$ WDs have mass
0.17 \msun\ and absolute magnitude $M_g=8.0$ \citep[see][]{kilic11a,
vennes11}.  We summarize the observed and derived stellar parameters of the 17 new
binaries in Table \ref{tab:param}.  Position and de-reddened $g$-band magnitude 
come from SDSS \citep{aihara11} and $d_{helio}$ is our heliocentric distance 
estimate.

\begin{figure*}          
 \plottwo{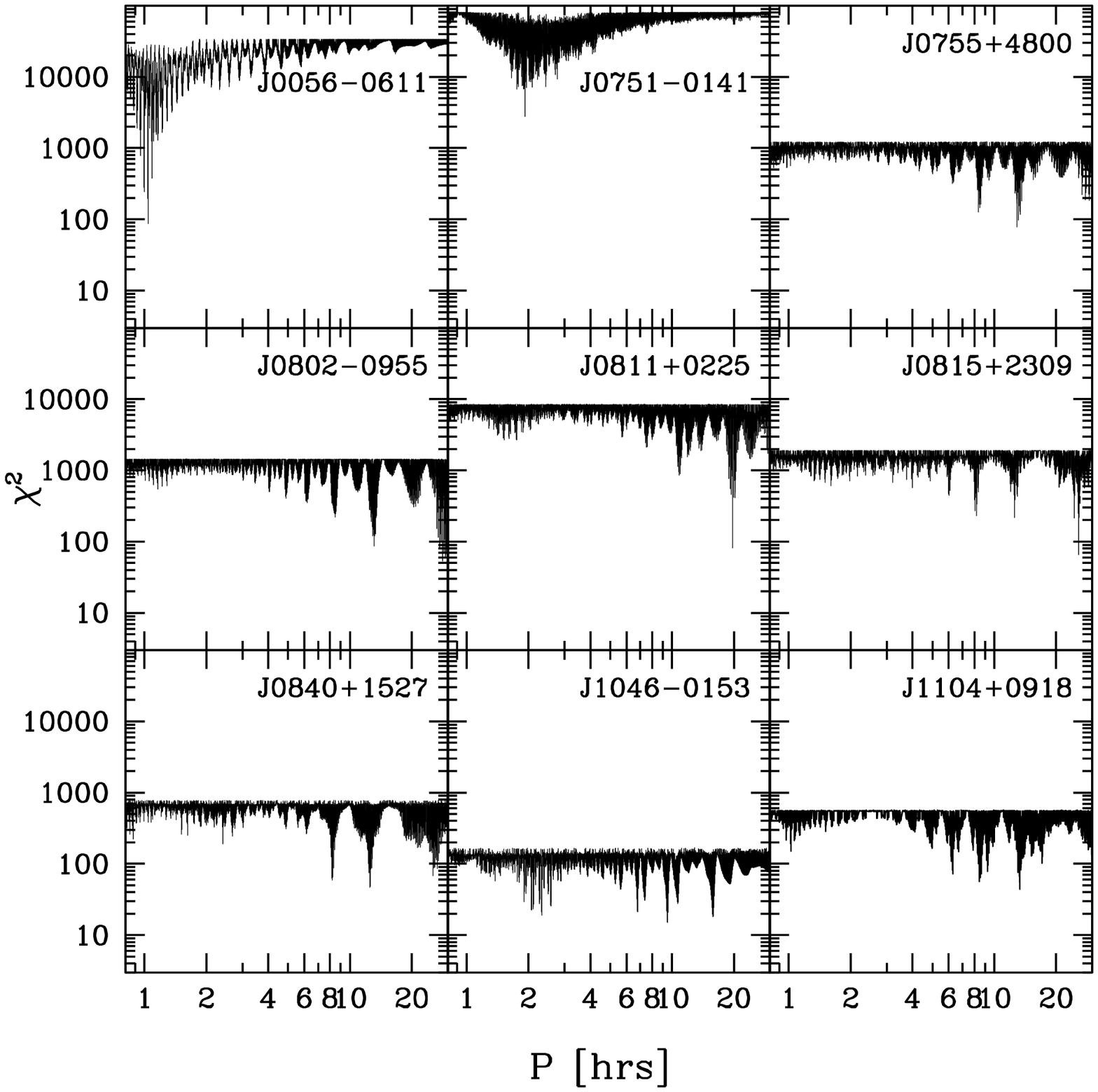}{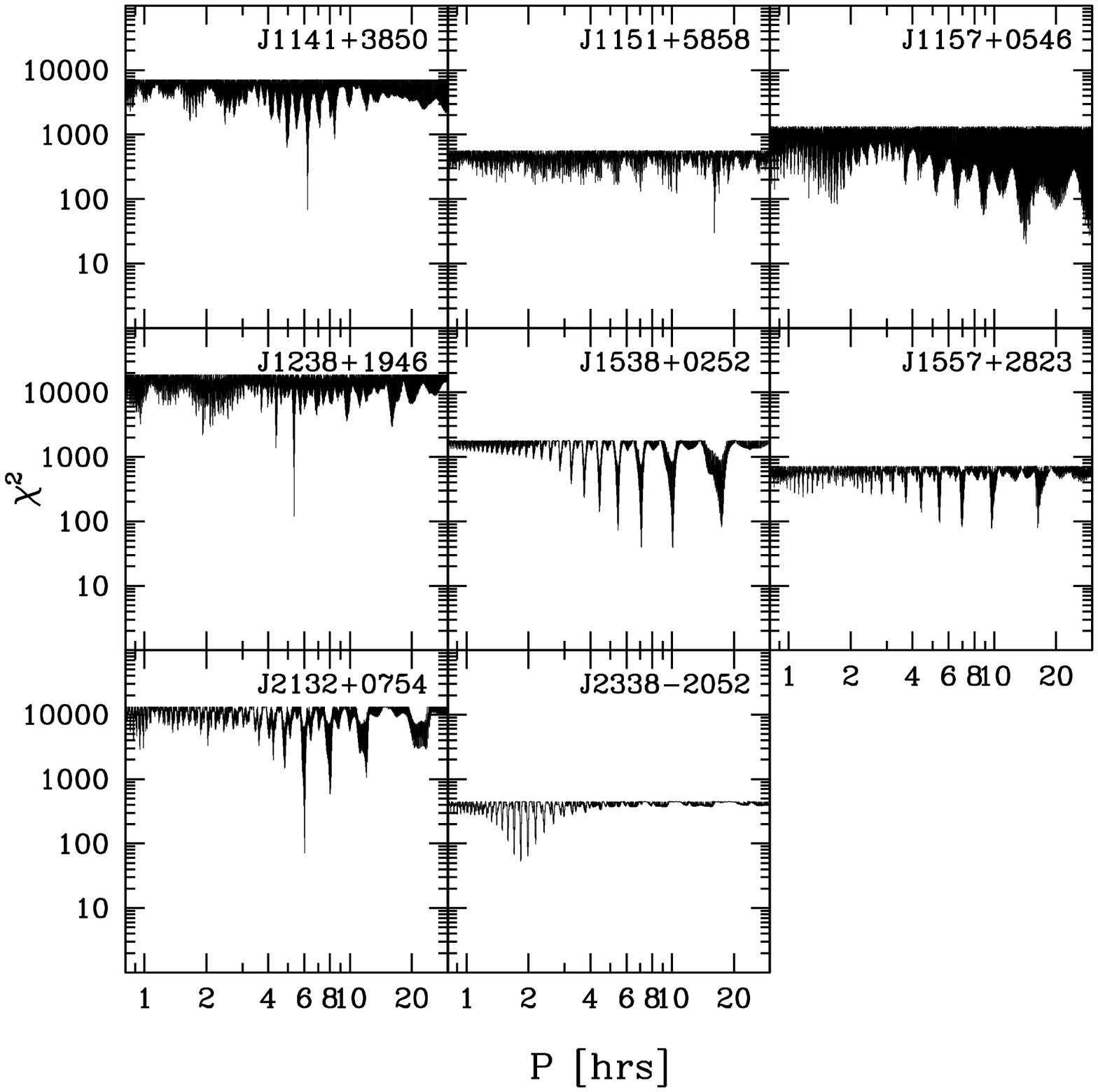}
 \caption{ \label{fig:pdm}
        Periodograms for the 17 WD binaries.  The best orbital periods have the 
smallest $\chi^2$ values; some binaries are well constrained and some have period 
aliases. }
 \end{figure*}

\begin{figure*}          
 \plottwo{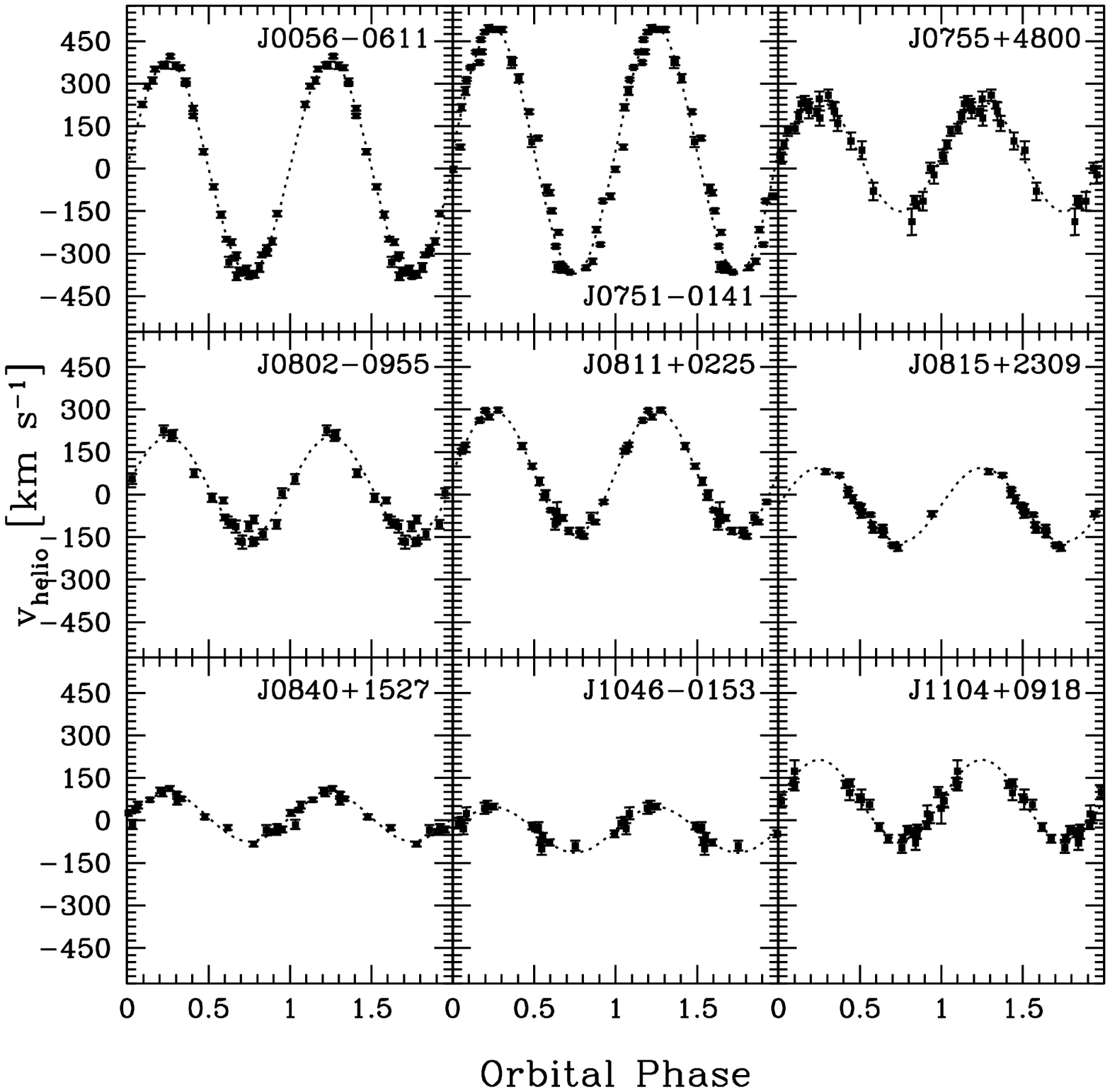}{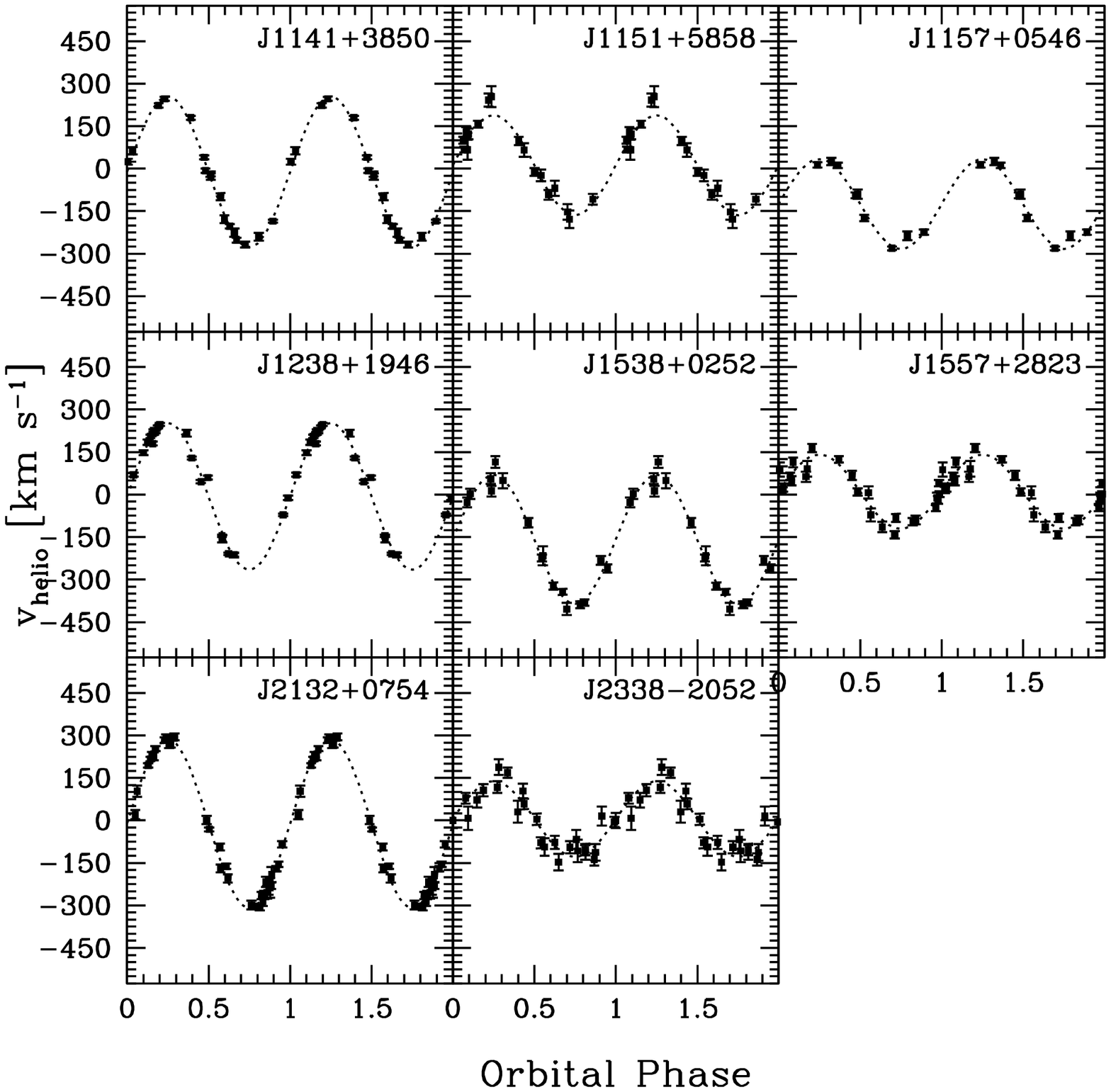}
 \caption{ \label{fig:vel}
        Observed velocities phased to best-fit orbits for the 17 WD binaries (Table
\ref{tab:orbit}). }
 \end{figure*}


\subsection{Orbital Elements}

	We calculate orbital elements and merger times in the same way as previous
ELM Survey papers, and so we refer the reader to those papers for the details of our
analysis.  In brief, we measure absolute radial velocities using the
cross-correlation package RVSAO \citep{kurtz98} and a high-S/N template.  We use the
entire spectrum in the cross-correlation.  We then use the summed spectra (Figure
\ref{fig:spec}) as cross-correlation templates to maximize our velocity precision
for each individual object.  We calculate orbital elements by minimizing $\chi^2$
for a circular orbit using the code of \citet{kenyon86}.  Figure \ref{fig:pdm} shows
the periodograms for the 17 binaries, and Figure \ref{fig:vel} plots the radial
velocities phased to the best-fit orbital periods.  We use the binary mass function
to estimate the unseen companion mass; an edge-on orbit with inclination
$i=90\arcdeg$ yields the minimum companion mass $M_2$ and the maximum gravitational
wave merger time.

        Table \ref{tab:orbit} summarizes the binary orbital parameters.  Columns
include orbital period ($P$), radial velocity semi-amplitude ($K$), systemic
velocity ($\gamma$), time of spectroscopic conjunction (the time when the object is
closest to us), minimum secondary mass ($M_2$) assuming $i=90\arcdeg$, the maximum
mass ratio ($q$), and the maximum gravitational wave merger time $\tau_{\rm merge}$.  
The systemic velocities in Table \ref{tab:orbit} are not corrected for the WDs'
gravitational redshifts, which should be subtracted from the observed velocities to
find the true systemic velocities. This correction is a few km s$^{-1}$ for a 0.17
\msun\ helium WD, comparable to the systemic velocity uncertainty.

\section{RESULTS}

        The orbital solutions constrain the nature of the ELM WD binaries.  Here we
discuss the systems with short merger times, massive companions, or that may be
underluminous supernovae progenitors.

\subsection{J0056$-$0611}

        The ELM WD J0056$-$0611 has a well-constrained orbital period of
$1.0409\pm0.0005$ hr and a semi-amplitude of $377\pm2$ \kms.  We can calculate its
likely companion mass if we assume a distribution for the unknown orbital
inclination.  Although radial velocity detections are biased towards edge-on systems
(see Section 4), we will assume that we are observing a random inclination for
purpose of discussion.  The mean inclination angle for a random sample,
$i=60\arcdeg$, is then an estimate of the most likely companion mass.  For
J0056$-$0611, the most likely companion is a 0.61 \msun\ WD at an orbital separation
of 0.5 \rsun.  This orbital separation rules out the possibility of a main sequence
companion.

	There is no evidence for a 0.61 \msun\ WD in the spectrum of J0056$-$0611,
but we would not expect there to be.  If the two WDs in this binary formed at the
same time, we would expect the 0.61 \msun\ WD to be 15 - 100 times less luminous
than the 0.17 \msun\ WD for cooling ages of 100 Myr - 1 Gyr \citep{bergeron95}.  A
more plausible evolutionary scenario for an ELM WD binary like J0056$-$0611 is two
consecutive phases of common-envelope evolution in which the ELM WD is created last,
giving the more massive WD a longer time to cool and fade \citep[e.g.][]{kilic07b}.

        For the most probable companion mass of 0.61 \msun, J0056$-$0611 will
begin mass transfer in 100 Myr.  \citet{kilic10} discuss the many possible stellar
evolution paths for such a system.  This system's mass ratio $q\leq$0.37 suggests
that mass transfer will likely be stable \citep{marsh04} and that J0056$-$0611
will evolve into an AM CVn system.

\subsection{J0751$-$0141}

        The ELM WD J0751$-$0141 has a $1.920\pm0.067$ hr orbital period with aliases
ranging between 1.85 and 2.05 hr (see Figure \ref{fig:pdm}).  The large $433\pm2$
\kms\ semi-amplitude indicates that the companion is massive, regardless of the
exact period.  The minimum companion mass is 0.94 \msun.  Assuming a random
inclination distribution, there is a 47\% probability that the companion is $>$1.4
\msun.

        Given that the ELM WD went through a common envelope phase of evolution with
its companion, J0751$-$0141 possibly contains a milli-second pulsar binary
companion.  Helium-core WDs are the most common type of companion in known
milli-second pulsar binaries \citep{tauris12}.  We have been allocated Cycle 14 {\it
Chandra X-ray Observatory} time to search for X-ray emission from a possible neutron
star.  

        For $i=60\arcdeg$, the companion is a 1.32 \msun\ WD, and the system will 
begin mass transfer in 290 Myr.  The extreme mass ratio $q\leq$0.18 means that this 
system will undergo stable mass transfer and evolve into an AM CVn system. As the 
binary orbit widens in the AM CVn phase, the mass accretion rate will drop and the 
mass required for the unstable burning of the accreted He-layer increases up to 
several percent of a solar mass.  The final flash should ignite a thermonuclear 
transient visible as an underluminous supernova \citep{bildsten07,shen09}. It is 
also possible that the helium flash will detonate the massive WD in a 
double-detonation scenario \citep[][though see Dan et al.\ 2012]{sim12}. If 
J0751$-$0141 has a massive WD companion, it is a probable supernova progenitor.

\subsection{J0811+0225}

        The ELM WD J0811+0225 has an orbital period of $19.727\pm0.012$ hr and a 
semi-amplitude of 221 \kms.  These orbital parameters yield a minimum companion mass 
of 1.20 \msun.  That means the companion probably exceeds a Chandrasekhar mass.  
For $i=60\arcdeg$, the most likely companion is a 1.70 \msun\ neutron star at an 
orbital separation of 4.6 \rsun.  There is no Fermi gamma-ray detection at this 
location, but additional observations are needed to determine the nature of this 
system.

\subsection{J0840+1527}

        The ELM WD J0840+1527 has a best-fit \logg $=5.043\pm0.053$ near the limit 
of our model grid.  Its best-fit orbital period is $12.517\pm0.114$ hr with a 
significant alias at 8.3 hr.  Assuming this object is 0.17 \msun, its most likely 
companion is a WD with a comparable mass, 0.19 \msun, at an orbital separation of 
1.9 \rsun.  If, on the other hand, J0840+1527 were a 3 \msun\ main sequence star, 
its companion would have an orbital separation of 4.3 \rsun\ -- a separation 
comparable to the radius of a 3 \msun\ star, and thus physically implausible.  
There is no evidence for mass transfer in this system.  We conclude that J0840+1527 
is a pair of ELM WDs.

\subsection{J1141+3850, J1157+0546, and J1238+1946}

	J1141+3850, J1157+0546, and J1238+1946 are the other systems containing WDs
near the low-gravity limit of our stellar atmosphere model grid, but their $k=158$ -
266 \kms\ semi-amplitudes are significantly larger than that of J0840+1527.  If we
assume that the objects are 0.17 \msun\ ELM WDs, then the binary companions have
minimum masses of 0.45 - 0.75 \msun.  If we instead assume that the objects are main
sequence stars, then the required orbital separations are comparable to the radius
of the main sequence star and physically impossible.  There is no evidence for mass
transfer in these systems.

	We conclude that J1141+3850, J1157+0546, and J1238+1946 are ELM WDs with 
likely WD companions.  For J1141+3850, there is a 36\% probability that the 
companion is $>$1.4 \msun, possibly a milli-second pulsar.  For J1141+3850 and 
J1238+1946, mass transfer will begin in 7-10 Gyr, making them AM CVn progenitors and 
possible underluminous supernovae progenitors.

\subsection{J2132+0754}

        The ELM WD J2132+0754 has a well-constrained orbital period of
$6.0134\pm0.0004$ hr and semi-amplitude of $297\pm3$ \kms.  The minimum companion
mass is 0.95 \msun, and there is a 48\% probability that the companion is $>$1.4
\msun, possibly a milli-second pulsar.  For $i=60\arcdeg$, the most likely companion
is a 1.33 \msun\ WD that will begin mass transfer in 6 Gyr.  That makes J2132+0754 a
likely AM CVn progenitor and another possible underluminous supernovae progenitor.

\subsection{J2338$-$2052}

        The ELM WD J2338$-$2052 has an orbital period of $1.834\pm0.170$ hr and a
semi-amplitude of $133\pm7$ \kms.  In this case the companion is another ELM WD; for
$i=60\arcdeg$, the most likely companion is a 0.17 \msun\ WD at an orbital
separation of 0.56 \rsun.  Given the unity mass ratio, this system will undergo
unstable mass transfer and will merge to form a single $\sim$0.4 \msun\ WD. This
system will merge in less than 1 Gyr.

\begin{figure}          
 \plotone{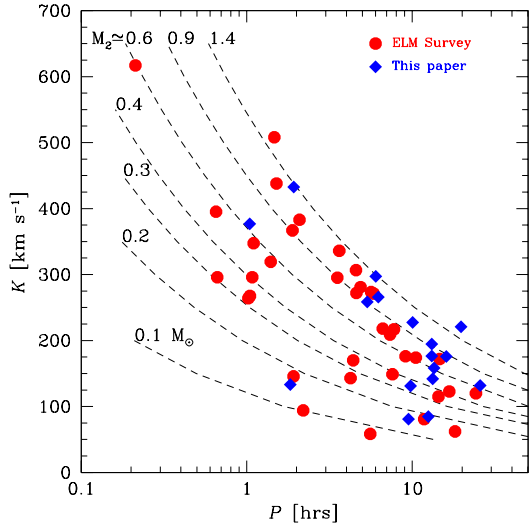}
 \caption{ \label{fig:kp}
        Orbital period vs.\ semi-amplitude for the ELM Survey binaries.  Previously
published binaries are drawn with solid red circles; the 17 binaries from this paper
are drawn with solid blue diamonds.  Dashed lines indicate approximate companion
mass under the assumption that $M_1=0.2$ \msun\ and $i=60\arcdeg$. }
 \end{figure}

\section{DISCUSSION}

        With these 17 new discoveries, plus the pulsating ELM WD discovery published
by \citet{hermes12d}, the ELM Survey has found 54 detached, double degenerate
binaries; 31 of the binaries will merge within a Hubble time.  Table
\ref{tab:summary} summarizes the properties of the systems.  Eighty percent of the
ELM Survey binaries are formally ELM WD systems.

\subsection{Significance of Binary Detections}

        We find low mass WDs in compact binaries, binaries that must have gone
through common envelope evolution.  This makes sense because extremely low mass WDs
require significant mass loss to form; the Universe is not old enough to produce an
extremely low mass WD through single star evolution.
        Yet four objects published in the ELM Survey have no significant velocity
variability, two of which are ELM WDs (see Table \ref{tab:summary}).  To understand
whether or not these stars are single requires that we understand the significance
of our binary detections.

        Each ELM Survey binary is typically constrained by 10-30 irregularly
spaced velocities with modest errors.  Given that we determine orbital parameters by
minimizing $\chi^2$, the $F$-test is a natural choice.  We use the $F$-test to check
whether the variance of the data around the orbital fit is consistent with the
variance of the data around a constant velocity (we use the weighted mean of the
observations).  $F$-test probabilities for the published ELM Survey binaries are
$<0.01$.  In other words, our binaries have significant velocity variability at the
$>99$\% confidence level.

        In the null cases we need to calculate the likelihood of not detecting a
binary.  This is a trickier problem, and one that we approach with a Monte Carlo
calculation.  We start by selecting a set of observations (times, velocity errors)
and an orbital period and semi-amplitude.  We convert observation times to orbital
phases using a randomly drawn zero time, and calculate velocities at those phases
summed with a randomly drawn velocity error.  We perform the $F$-test, using 0.01 as
a detection threshold.  We repeat this calculation 10,000 times for a given orbital
period and semi-amplitude, and then select a new orbital period and semi-amplitude
to iterate on.  This analysis is done for each object.

        We find that the datasets for our 17 new binaries have a median 99.9\%
likelihood of detecting $K=200$ \kms\ binaries, a 97\% likelihood of detecting
$K=100$ \kms\ binaries, and a 44\% likelihood of detecting $K=50$ \kms\ binaries. It
is no surprise that we are less likely to detect a low semi-amplitude binary, but
this analysis suggests that we can be quite confident of detecting $K>100$ \kms\
systems.  We find very similar likelihoods for detecting binaries containing ELM WDs
in the full ELM Survey sample.

        The datasets for the null cases typically contain fewer observations and so
are not as well-constrained.  For J0900+0234, a 0.16 \msun\ ELM WD with no observed
velocity variation \citep{brown12a}, the likelihoods of detecting a $K=200$, 100,
and 50 \kms\ binary are 87\%, 57\%, and 10\%, respectively.  Additional observations
are required to claim this ELM WD as non-variable.

        There is, of course, an orbital period dependence to the detections, and
periods near 24 hr are the most problematic.  Taken together, our datasets have a
median 39\% likelihood of detecting a $K=100$ \kms\ binary at $P=24$ hr.  Yet we
remain sensitive to longer periods:  our datasets have median 99\% and 98\%
likelihoods of detecting a $K=100$ \kms\ binary at $P=18$ hr and $P=36$ hr,
respectively.

        Figure \ref{fig:kp} plots the observed distribution of $P$ and $K$ for the
ELM Survey binaries.  The dashed lines indicate the approximate companion mass
assuming $M_1$ is 0.2 \msun\ and $i$ is 60$\arcdeg$.  At $K=100$ \kms, we can
detect companion masses down to 0.1 \msun\ at 2 hr orbital periods and 0.55 \msun\
companions at 2 day orbital periods.  Systems with $K<100$ \kms\ are the realm of
$\lesssim$0.2 \msun\ companions, and we observe a half-dozen systems with these
parameters.  Our incompleteness at $K<100$ \kms\ implies there are quite likely more
ELM WDs with $\simeq$0.2 \msun\ companions; the remaining ELM WD candidates that
do not show obvious velocity change in a couple observations are possible examples
of such low amplitude systems.

        Finally, orbital inclination acts to increase the difficulty of 
identifying lone ELM WDs.  Consider the set of 46 ELM WDs in Table 
\ref{tab:summary} with $\log{g}<7$, two of which are non-variable.  Their 
median $K$ is 240 \kms, a semi-amplitude that would appear $<50$ \kms\ at 
$i<12\arcdeg$.  If the 46 objects have randomly distributed orbital inclinations, 
one object should have $i<12\arcdeg$ and a second $i<17\arcdeg$.  We conclude 
there is no good evidence for a lone ELM WD in our present sample. This is in 
stark contrast to the population of 0.4 $M_{\odot}$ WDs in the solar neighborhood, 
of which 20\%-30\% are single \citep{brown11}.

\begin{figure}          
 \plotone{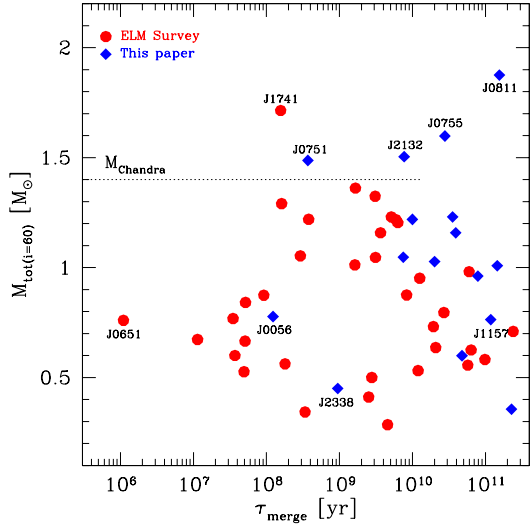}
 \caption{ \label{fig:merge}
        Gravitational wave merger time versus total system mass for the ELM Survey.
We calculate system mass assuming $i=60^{\arcdeg}$ when orbital inclination is
unknown.
        Previously published ELM Survey binaries are drawn with solid red circles, 
and the 17 new binaries from this paper are drawn with solid blue diamonds.  Six 
binaries have probable masses exceeding the Chandrasekhar mass; three have merger 
times less than 10 Gyr, as indicated by the dotted line.  The minimum companion mass 
for J0811+0225 is 1.2 $M_{\odot}$.}
 \end{figure}

\begin{figure}          
 \plotone{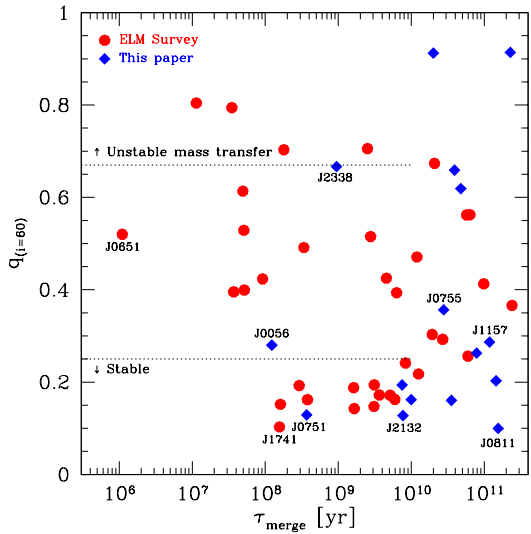}
 \caption{ \label{fig:tau}
        Gravitational wave merger time versus mass ratio $q$ for the ELM Survey. 
We calculate mass ratio assuming $i=60^{\arcdeg}$ when orbital 
inclination is unknown.
        Previously published ELM Survey binaries are drawn with solid red circles, 
and the 17 new binaries from this paper are drawn with solid blue diamonds.  Dotted 
lines mark the approximate thresholds for stability of mass transfer during the 
Roche lobe overflow phase \citep{marsh04}, drawn for systems that will merge in 
$<$10 Gyr. Stable mass transfer systems will evolve into AM CVn systems; unstable 
mass transfer systems will possibly merge.}
 \end{figure}

\subsection{The Future of ELM WDs}

        One of the most exciting aspects of our ELM WD binaries is that many have
gravitational wave merger times less than a Hubble time, and in one case as short as
1 Myr \citep{brown11b, hermes12c}.  A natural question, then, is what will happen
when the WDs merge.  Notably, 5 of our 17 new binaries have probable system masses
(for $i=60\arcdeg$) in excess of a Chandrasekhar mass.

        Figure \ref{fig:merge} plots the distribution of merger time and system mass
for the ELM Survey binaries assuming $i=60\arcdeg$, except when inclination is
known.  The majority of our binaries have probable masses below the Chandrasekhar
mass and are not potential supernova Type Ia progenitors, but six systems appear to
have either neutron star or massive WD companions.  The evolution of these six
systems depends on the stability of mass transfer during the Roche lobe overflow
phase, and thus on the mass ratio of the binary.

        Figure \ref{fig:tau} plots the distribution of merger time and mass ratio
$q$ for the ELM Survey binaries, again assuming $i=60\arcdeg$ except when
inclination is known.  The Chandrasekhar mass systems generally have extreme
$q\lesssim0.15$ and so will evolve into stable mass transfer AM CVn systems.  If
the accretors are massive WDs, these systems should experience a large helium
flash that may appear as an underluminous .Ia supernova \citep[e.g.][]{bildsten07,
shen09}.  It is also possible that the helium flash will detonate the massive WD
in a double-detonation scenario \citep{sim12}. The Chandrasekhar mass systems are
thus potential supernovae progenitors. However, the outcome of a merger
of an ELM WD with a massive WD is uncertain \citep[see][]{dan12}.

        Systems with $q\ge2/3$, such as J2338$-$2052, will experience unstable mass
transfer.  These systems may merge \citep{dan11} to form a single low-mass WD, R CrB
star, or helium-burning sdB star.  Systems with intermediate $q\simeq1/2$, such as
J0651+2844, will experience either stable or unstable mass transfer depending on
their spin-orbit coupling.
        After we constrain the remaining ELM WD candidates plotted in Fig.\
\ref{fig:teff} we look forward to calculating their space density and merger rate.
ELM WDs binaries are clearly an important channel for AM CVn star formation, and
thus an important source of strong gravitational waves in the mHz regime.

\section{CONCLUSION}

        We perform stellar atmosphere fits to the entire collection of single-epoch
spectra in the HVS Survey and identify 57 new low mass WD candidates.  Follow-up
spectroscopy reveals 17 WDs with significant velocity variability, 12 of which are
ELM WDs.  Presently, the ELM Survey sample is consistent with all of the ELM WDs
being part of close binaries.  ELM WDs are thus signposts for binaries that are
strong gravitational wave sources and possible supernovae progenitors.

        Four binaries in our sample contain $\log{g}\simeq5$ objects, objects that
might not be considered WDs or proto-WDs if not for their observed orbital motion.  
The existence of $\log{g}\simeq5$ WDs motivates us to develop a new set of stellar
atmosphere models following \citet{gianninas11}.  Interestingly, all of the lowest
surface gravity ($\log{g}<6$) WDs in our sample have metal lines in their spectra.
This issue will be studied in detail in a future paper; infrared observations are
needed to constrain the possibility of debris disks as the source of Ca and Mg
accretion in these compact binary systems.

	Recent discoveries of pulsations in several of our ELM WD targets
\citep{hermes12b, hermes12d} enable us to probe the interiors of low-mass,
presumably He-core WDs using the tools of asteroseismology. Due to the HVS color
selection, all of the newly identified systems are hotter than 10,000 K and
therefore too hot to show the g-mode pulsations detected in cooler ELM WDs
\citep{corsico12}.

        Four binaries in our sample have massive $\gtrsim$0.9 \msun\ companions and
$<$10 Gyr merger times.  If the unseen companions are massive WDs, these extreme
mass ratio binaries will undergo stable mass transfer and will evolve into AM CVn
systems and potentially future .Ia or underluminous supernovae. Thus the systems
J0751$-$0141, J1141+3850, J1238+1946, and J2132+0754 are possible supernova
progenitors.

	The other possibility is that the massive companions in these four binaries
are neutron stars.  Given their past common envelope evolution, the systems quite
possibly contain milli-second pulsars.  A fifth system, J0811+0225, appears to have
a minimum companion mass close to a Chandrasekhar mass.  We are conducting follow-up
optical, radio, and X-ray observations to establish the nature of these extreme mass
ratio ELM WD binaries. Optically visible WD companions of neutron stars are useful
for constraining the binary mass ratio and also to calibrate the spin-down ages of
milli-second pulsars.

	Given our present observations, we expect that the ELM Survey will grow to 
100 binaries in a few years.  A large sample is important for finding more systems 
like J0651+2844, a 12.75 min period system that provides us with a laboratory for 
measuring the spin-orbit coupling and tidal heating of a rapidly merging pair of 
WDs.  We expect that photometric observations will discover more pulsating ELM WDs.  
Finally, a larger sample of ELM WDs will be an important source of gravitational 
wave verification sources for $eLISA$ and other future gravitational wave detection 
experiments.

\acknowledgments
        We thank M.\ Alegria, P.\ Canton, S.\ Gotilla, J.\ McAfee, E.\ Martin, A.\
Milone, and R.\ Ortiz for their assistance with observations obtained at the MMT
Observatory, and P.\ Berlind and M.\ Calkins for their assistance with observations
obtained at the Fred Lawrence Whipple Observatory.  We also thank P.\ Bergeron and
D.\ Koester for their invaluable assistance in the computation of our stellar
atmosphere model grids.  This research makes use of the SAO/NASA Astrophysics Data
System Bibliographic Service.  This project makes use of data products from the
Sloan Digital Sky Survey, which has been funded by the Alfred P.\ Sloan Foundation,
the Participating Institutions, the National Science Foundation, and the U.S.\
Department of Energy Office of Science.  This work was supported in part by the
Smithsonian Institution.

{\it Facilities:} \facility{MMT (Blue Channel Spectrograph), FLWO:1.5m (FAST)}

\begin{deluxetable*}{crcccccccc}
\tabletypesize{\scriptsize}
\tablecolumns{10}
\tablewidth{0pt}
\tablecaption{Merger and Non-Merger Systems in the ELM Survey\label{tab:summary}}
\tablehead{
        \colhead{Object}&
        \colhead{\teff}&
        \colhead{$\log{g}$}&
        \colhead{$P$}&
        \colhead{$K$}&
        \colhead{Mass}&
        \colhead{$M_2$}&
        \colhead{$M_2(60\arcdeg)$}&
        \colhead{$\tau_{\rm merge}$}&
        \colhead{Ref}\\
  & (K) & (cm s$^{-2}$) & (days) & \kms & \msun & \msun & \msun & Gyr &
}
        \startdata
J0022$-$1014 & 18980  & 7.15  & 0.07989 & 145.6 & 0.33 & $\ge 0.19$ & 0.23    & $\le 0.73$  & 6 \\
J0056$-$0611 & 12210  & 6.17  & 0.04338 & 376.9 & 0.17 & $\ge 0.46$ & 0.61    & $\le 0.12$  &   \\
J0106$-$1000 & 16490  & 6.01  & 0.02715 & 395.2 & 0.17 &       0.43 & \nodata &     0.037   & 7 \\
J0112+1835   &  9690  & 5.63  & 0.14698 & 295.3 & 0.16 & $\ge 0.62$ & 0.85    & $\le 2.7$   & 1 \\
J0651+2844   & 16530  & 6.76  & 0.00886 & 616.9 & 0.26 &       0.50 & \nodata &    0.0011   & 3,15 \\
J0751$-$0141 & 15660  & 5.43  & 0.08001 & 432.6 & 0.17 & $\ge 0.94$ & 1.32    & $\le 0.37$  &   \\
J0755+4906   & 13160  & 5.84  & 0.06302 & 438.0 & 0.17 & $\ge 0.81$ & 1.12    & $\le 0.22$  & 2 \\
J0818+3536   & 10620  & 5.69  & 0.18315 & 170.0 & 0.17 & $\ge 0.26$ & 0.33    & $\le 8.9$   & 2 \\
J0822+2753   & 8880   & 6.44  & 0.24400 & 271.1 & 0.17 & $\ge 0.76$ & 1.05    & $\le 8.4$   & 4 \\
J0825+1152   & 24830  & 6.61  & 0.05819 & 319.4 & 0.26 & $\ge 0.47$ & 0.61    & $\le 0.18$  & 0 \\
J0849+0445   & 10290  & 6.23  & 0.07870 & 366.9 & 0.17 & $\ge 0.64$ & 0.88    & $\le 0.47$  & 4 \\
J0923+3028   & 18350  & 6.63  & 0.04495 & 296.0 & 0.23 & $\ge 0.34$ & 0.44    & $\le 0.13$  & 2 \\
J1005+0542   & 15740  & 7.25  & 0.30560 & 208.9 & 0.34 & $\ge 0.66$ & 0.86    & $\le 9.0$   & 0 \\
J1005+3550   & 10010  & 5.82  & 0.17652 & 143.0 & 0.17 & $\ge 0.19$ & 0.24    & $\le 10.3$  & 0 \\
J1053+5200   & 15180  & 6.55  & 0.04256 & 264.0 & 0.20 & $\ge 0.26$ & 0.33    & $\le 0.16$  & 4,9 \\
J1056+6536   & 20470  & 7.13  & 0.04351 & 267.5 & 0.34 & $\ge 0.34$ & 0.43    & $\le 0.085$ & 0 \\
J1112+1117   &  9590  & 6.36  & 0.17248 & 116.2 & 0.17 & $\ge 0.14$ & 0.17    & $\le 12.7$  & 16 \\
J1141$+$3850 & 11620  & 5.31  & 0.25958 & 265.8 & 0.17 & $\ge 0.76$ & 1.05    & $\le 9.96$  &   \\
J1233+1602   & 10920  & 5.12  & 0.15090 & 336.0 & 0.17 & $\ge 0.86$ & 1.20    & $\le 2.1$   & 2 \\
J1234$-$0228 & 18000  & 6.64  & 0.09143 & 94.0  & 0.23 & $\ge 0.09$ & 0.11    & $\le 2.7$   & 6 \\
J1238$+$1946 & 16170  & 5.28  & 0.22275 & 258.6 & 0.17 & $\ge 0.64$ & 0.88    & $\le 7.49$  &   \\
J1436+5010   & 16550  & 6.69  & 0.04580 & 347.4 & 0.24 & $\ge 0.46$ & 0.60    & $\le 0.10$  & 4,9 \\
J1443+1509   &  8810  & 6.32  & 0.19053 & 306.7 & 0.17 & $\ge 0.83$ & 1.15    & $\le 4.1$   & 1 \\
J1630+4233   & 14670  & 7.05  & 0.02766 & 295.9 & 0.30 & $\ge 0.30$ & 0.37    & $\le 0.031$ & 8 \\
J1741+6526   &  9790  & 5.19  & 0.06111 & 508.0 & 0.16 & $\ge 1.10$ & 1.55    & $\le 0.17$  & 1 \\
J1840+6423   &  9140  & 6.16  & 0.19130 & 272.0 & 0.17 & $\ge 0.64$ & 0.88    & $\le 5.0$   & 1 \\
J2103$-$0027 & 10000  & 5.49  & 0.20308 & 281.0 & 0.17 & $\ge 0.71$ & 0.99    & $\le 5.4$   & 0 \\
J2119$-$0018 & 10360  & 5.36  & 0.08677 & 383.0 & 0.17 & $\ge 0.75$ & 1.04    & $\le 0.54$  & 2 \\
J2132$+$0754 & 13700  & 6.00  & 0.25056 & 297.3 & 0.17 & $\ge 0.95$ & 1.33    & $\le 7.70$  &   \\
J2338$-$2052 & 16630  & 6.87  & 0.07644 & 133.4 & 0.27 & $\ge 0.15$ & 0.18    & $\le 0.95$  &   \\
NLTT11748    & 8690   & 6.54  & 0.23503 & 273.4 & 0.18 &      0.76  & \nodata &      7.2    & 5,10,11 \\
\tableline
J0022+0031   & 17890  & 7.38  & 0.49135 &  80.8 & 0.38 & $\ge 0.21$ & 0.26    & \nodata     & 6 \\
J0152+0749   & 10840  & 5.80  & 0.32288 & 217.0 & 0.17 & $\ge 0.57$ & 0.78    & \nodata     & 1 \\
J0730+1703   & 11080  & 6.36  & 0.69770 & 122.8 & 0.17 & $\ge 0.32$ & 0.41    & \nodata     & 0 \\
J0755$+$4800 & 19890  & 7.46  & 0.54627 & 194.5 & 0.42 & $\ge 0.90$ & 1.18    & \nodata     &   \\
J0802$-$0955 & 16910  & 6.42  & 0.54687 & 176.5 & 0.20 & $\ge 0.57$ & 0.76    & \nodata     &   \\
J0811$+$0225 & 13990  & 5.79  & 0.82194 & 220.7 & 0.17 & $\ge 1.20$ & 1.71    & \nodata     &   \\
J0815$+$2309 & 21470  & 5.78  & 1.07357 & 131.7 & 0.17 & $\ge 0.47$ & 0.63    & \nodata     &   \\
J0840$+$1527 & 13810  & 5.04  & 0.52155 &  84.8 & 0.17 & $\ge 0.15$ & 0.19    & \nodata     &   \\
J0845+1624   & 17750  & 7.42  & 0.75599 &  62.2 & 0.40 & $\ge 0.19$ & 0.22    & \nodata     & 0 \\
J0900+0234   &  8220  & 5.78  & \nodata & $\le24$ & 0.16 & \nodata  & \nodata & \nodata     & 1 \\
J0917+4638   & 11850  & 5.55  & 0.31642 & 148.8 & 0.17 & $\ge 0.28$ & 0.36    & \nodata     & 12 \\
J1046$-$0153 & 14880  & 7.37  & 0.39539 &  80.8 & 0.37 & $\ge 0.19$ & 0.23    & \nodata     &   \\
J1104$+$0918 & 16710  & 7.61  & 0.55319 & 142.1 & 0.46 & $\ge 0.55$ & 0.70    & \nodata     &   \\
J1151$+$5858 & 15400  & 6.09  & 0.66902 & 175.7 & 0.17 & $\ge 0.61$ & 0.84    & \nodata     &   \\
J1157$+$0546 & 12100  & 5.05  & 0.56500 & 158.3 & 0.17 & $\ge 0.45$ & 0.59    & \nodata     &   \\
J1422+4352   & 12690  & 5.91  & 0.37930 & 176.0 & 0.17 & $\ge 0.41$ & 0.55    & \nodata     & 2 \\
J1439+1002   & 14340  & 6.20  & 0.43741 & 174.0 & 0.18 & $\ge 0.46$ & 0.62    & \nodata     & 2 \\
J1448+1342   & 12580  & 6.91  & \nodata & $\le 35$ & 0.25 & \nodata    & \nodata & \nodata     & 2 \\
J1512+2615   & 12130  & 6.62  & 0.59999 & 115.0 & 0.20 & $\ge 0.28$ & 0.36    & \nodata     & 2 \\
J1518+0658   &  9810  & 6.66  & 0.60935 & 172.0 & 0.20 & $\ge 0.58$ & 0.78    & \nodata     & 1 \\
J1538$+$0252 & 11560  & 5.97  & 0.41915 & 227.6 & 0.17 & $\ge 0.77$ & 1.06    & \nodata     &   \\
J1557$+$2823 & 12550  & 7.76  & 0.40741 & 131.2 & 0.49 & $\ge 0.43$ & 0.54    & \nodata     &   \\
J1625+3632   & 23570  & 6.12  & 0.23238 &  58.4 & 0.20 & $\ge 0.07$ & 0.08    & \nodata     & 6 \\
J1630+2712   & 11200  & 5.95  & 0.27646 & 218.0 & 0.17 & $\ge 0.52$ & 0.70    & \nodata     & 2 \\
J2252$-$0056 & 19450  & 7.00  & \nodata & $\le25$ & 0.31 & \nodata  & \nodata & \nodata     & 2 \\
J2345$-$0102 & 33130  & 7.20  & \nodata & $\le43$ & 0.42 & \nodata  & \nodata & \nodata     & 2 \\
LP400$-$22   & 11170  & 6.35  & 1.01016 & 119.9 & 0.19 & $\ge 0.41$ & 0.52    & \nodata     & 13,14
        \enddata
\tablerefs{ (0) \citet{kilic12a}; (1) \citet{brown12a}; (2) \citet{brown10c};
        (3) \citet{brown11b}; (4) \citet{kilic10}; (5) \citet{kilic10b};
        (6) \citet{kilic11a}; (7) \citet{kilic11b}; (8) \citet{kilic11c};
        (9) \citet{mullally09}; (10) \citet{steinfadt10}; (11) \citet{kawka10};
        (12) \citet{kilic07}; (13) \citet{kilic09}; (14) \citet{vennes09} ;
	(15) \citet{hermes12c} ; (16) \citet{hermes12d} }

\end{deluxetable*}

\clearpage

\begin{thebibliography}{73}
\expandafter\ifx\csname natexlab\endcsname\relax\def\natexlab#1{#1}\fi

\bibitem[{{Aihara} {et~al.}(2011){Aihara}, {Allende Prieto}, {An},
  {et~al.}}]{aihara11}
{Aihara}, H., {Allende Prieto}, C., {An}, D., {et~al.} 2011, \apjs, 193, 29

\bibitem[{{Allende Prieto} {et~al.}(2006){Allende Prieto}, {Beers}, {Wilhelm},
  {et~al.}}]{allende06}
{Allende Prieto}, C., {Beers}, T.~C., {Wilhelm}, R., {et~al.} 2006, \apj, 636,
  804

\bibitem[{{Allende Prieto} {et~al.}(2008){Allende Prieto}, {Sivarani}, {Beers},
  {et~al.}}]{allende08}
{Allende Prieto}, C., {Sivarani}, T., {Beers}, T.~C., {et~al.} 2008, \aj, 136,
  2070

\bibitem[{{Bassa} {et~al.}(2006){Bassa}, {van Kerkwijk}, {Koester}, \&
  {Verbunt}}]{bassa06}
{Bassa}, C.~G., {van Kerkwijk}, M.~H., {Koester}, D., \& {Verbunt}, F. 2006,
  \aap, 456, 295

\bibitem[{{Bergeron} {et~al.}(1995){Bergeron}, {Wesemael}, \&
  {Beauchamp}}]{bergeron95}
{Bergeron}, P., {Wesemael}, F., \& {Beauchamp}, A. 1995, \pasp, 107, 1047

\bibitem[{{Bildsten} {et~al.}(2007){Bildsten}, {Shen}, {Weinberg}, \&
  {Nelemans}}]{bildsten07}
{Bildsten}, L., {Shen}, K.~J., {Weinberg}, N.~N., \& {Nelemans}, G. 2007,
  \apjl, 662, L95

\bibitem[{{Brown} {et~al.}(2011{\natexlab{a}}){Brown}, {Kilic}, {Brown}, \&
  {Kenyon}}]{brown11}
{Brown}, J.~M., {Kilic}, M., {Brown}, W.~R., \& {Kenyon}, S.~J.
  2011{\natexlab{a}}, \apj, 729, 2

\bibitem[{{Brown} {et~al.}(2009){Brown}, {Geller}, \& {Kenyon}}]{brown09a}
{Brown}, W.~R., {Geller}, M.~J., \& {Kenyon}, S.~J. 2009, \apj, 690, 1639

\bibitem[{{Brown} {et~al.}(2012{\natexlab{a}}){Brown}, {Geller}, \&
  {Kenyon}}]{brown12b}
---. 2012{\natexlab{a}}, \apj, 751, 55

\bibitem[{{Brown} {et~al.}(2005){Brown}, {Geller}, {Kenyon}, \&
  {Kurtz}}]{brown05}
{Brown}, W.~R., {Geller}, M.~J., {Kenyon}, S.~J., \& {Kurtz}, M.~J. 2005,
  \apjl, 622, L33

\bibitem[{{Brown} {et~al.}(2006{\natexlab{a}}){Brown}, {Geller}, {Kenyon}, \&
  {Kurtz}}]{brown06}
---. 2006{\natexlab{a}}, \apjl, 640, L35

\bibitem[{{Brown} {et~al.}(2006{\natexlab{b}}){Brown}, {Geller}, {Kenyon}, \&
  {Kurtz}}]{brown06b}
---. 2006{\natexlab{b}}, \apj, 647, 303

\bibitem[{{Brown} {et~al.}(2007{\natexlab{a}}){Brown}, {Geller}, {Kenyon},
  {Kurtz}, \& {Bromley}}]{brown07a}
{Brown}, W.~R., {Geller}, M.~J., {Kenyon}, S.~J., {Kurtz}, M.~J., \& {Bromley},
  B.~C. 2007{\natexlab{a}}, \apj, 660, 311

\bibitem[{{Brown} {et~al.}(2007{\natexlab{b}}){Brown}, {Geller}, {Kenyon},
  {Kurtz}, \& {Bromley}}]{brown07b}
---. 2007{\natexlab{b}}, \apj, 671, 1708

\bibitem[{{Brown} {et~al.}(2010){Brown}, {Kilic}, {Allende Prieto}, \&
  {Kenyon}}]{brown10c}
{Brown}, W.~R., {Kilic}, M., {Allende Prieto}, C., \& {Kenyon}, S.~J. 2010,
  \apj, 723, 1072

\bibitem[{{Brown} {et~al.}(2011{\natexlab{b}}){Brown}, {Kilic}, {Allende
  Prieto}, \& {Kenyon}}]{brown11a}
---. 2011{\natexlab{b}}, \mnras, 411, L31

\bibitem[{{Brown} {et~al.}(2012{\natexlab{b}}){Brown}, {Kilic}, {Allende
  Prieto}, \& {Kenyon}}]{brown12a}
---. 2012{\natexlab{b}}, \apj, 744, 142

\bibitem[{{Brown} {et~al.}(2011{\natexlab{c}}){Brown}, {Kilic}, {Hermes},
  {et~al.}}]{brown11b}
{Brown}, W.~R., {Kilic}, M., {Hermes}, J.~J., {et~al.} 2011{\natexlab{c}},
  \apjl, 737, L23

\bibitem[{{Cocozza} {et~al.}(2006){Cocozza}, {Ferraro}, {Possenti}, \&
  {D'Amico}}]{cocozza06}
{Cocozza}, G., {Ferraro}, F.~R., {Possenti}, A., \& {D'Amico}, N. 2006, \apjl,
  641, L129

\bibitem[{{C{\'o}rsico} {et~al.}(2012){C{\'o}rsico}, {Romero}, {Althaus}, \&
  {Hermes}}]{corsico12}
{C{\'o}rsico}, A.~H., {Romero}, A.~D., {Althaus}, L.~G., \& {Hermes}, J.~J.
  2012, \aap, 547, A96

\bibitem[{{Dan} {et~al.}(2011){Dan}, {Rosswog}, {Guillochon}, \&
  {Ramirez-Ruiz}}]{dan11}
{Dan}, M., {Rosswog}, S., {Guillochon}, J., \& {Ramirez-Ruiz}, E. 2011, \apj,
  737, 89

\bibitem[{{Dan} {et~al.}(2012){Dan}, {Rosswog}, {Guillochon}, \&
  {Ramirez-Ruiz}}]{dan12}
---. 2012, \mnras, 422, 2417

\bibitem[{{Dorman} {et~al.}(1993){Dorman}, {Rood}, \& {O'Connell}}]{dorman93}
{Dorman}, B., {Rood}, R.~T., \& {O'Connell}, R.~W. 1993, \apj, 419, 596

\bibitem[{{Fabricant} {et~al.}(1998){Fabricant}, {Cheimets}, {Caldwell}, \&
  {Geary}}]{fabricant98}
{Fabricant}, D., {Cheimets}, P., {Caldwell}, N., \& {Geary}, J. 1998, \pasp,
  110, 79

\bibitem[{{Geier} {et~al.}(2012){Geier}, {Schaffenroth}, {Hirsch},
  {et~al.}}]{geier12}
{Geier}, S., {Schaffenroth}, V., {Hirsch}, H., {et~al.} 2012, in ASP Conf.
  Ser,, Vol. 452, Fifth Meeting on Hot Subdwarf Stars and Related Objects, ed.
  D.~{Kilkenny}, C.~S. {Jeffery}, \& C.~{Koen}, 129

\bibitem[{{Gianninas} {et~al.}(2011){Gianninas}, {Bergeron}, \&
  {Ruiz}}]{gianninas11}
{Gianninas}, A., {Bergeron}, P., \& {Ruiz}, M.~T. 2011, \apj, 743, 138

\bibitem[{{Girardi} {et~al.}(2002){Girardi}, {Bertelli}, {Bressan},
  {et~al.}}]{girardi02}
{Girardi}, L., {Bertelli}, G., {Bressan}, A., {et~al.} 2002, \aap, 391, 195

\bibitem[{{Girardi} {et~al.}(2004){Girardi}, {Grebel}, {Odenkirchen}, \&
  {Chiosi}}]{girardi04}
{Girardi}, L., {Grebel}, E.~K., {Odenkirchen}, M., \& {Chiosi}, C. 2004, \aap,
  422, 205

\bibitem[{{Heber}(2009)}]{heber09}
{Heber}, U. 2009, \araa, 47, 211

\bibitem[{{Heber} {et~al.}(2003){Heber}, {Edelmann}, {Lisker}, \&
  {Napiwotzki}}]{heber03}
{Heber}, U., {Edelmann}, H., {Lisker}, T., \& {Napiwotzki}, R. 2003, \aap, 411,
  L477

\bibitem[{{Hermes} {et~al.}(2012{\natexlab{a}}){Hermes}, {Kilic}, {Brown},
  {Montgomery}, \& {Winget}}]{hermes12a}
{Hermes}, J.~J., {Kilic}, M., {Brown}, W.~R., {Montgomery}, M.~H., \& {Winget},
  D.~E. 2012{\natexlab{a}}, \apj, 749, 42

\bibitem[{{Hermes} {et~al.}(2012{\natexlab{b}}){Hermes}, {Kilic}, {Brown},
  {et~al.}}]{hermes12c}
{Hermes}, J.~J., {Kilic}, M., {Brown}, W.~R., {et~al.} 2012{\natexlab{b}},
  \apjl, 757, L21

\bibitem[{{Hermes} {et~al.}(2012{\natexlab{c}}){Hermes}, {Montgomery},
  {Winget}, {Brown}, {Kilic}, \& {Kenyon}}]{hermes12b}
{Hermes}, J.~J., {Montgomery}, M.~H., {Winget}, D.~E., {Brown}, W.~R., {Kilic},
  M., \& {Kenyon}, S.~J. 2012{\natexlab{c}}, \apjl, 750, L28

\bibitem[{{Hermes} {et~al.}(2013){Hermes}, {Montgomery}, {Winget},
  {et~al.}}]{hermes12d}
{Hermes}, J.~J., {Montgomery}, M.~H., {Winget}, D.~E., {et~al.} 2013, \apj,
  accepted

\bibitem[{{Kaplan} {et~al.}(2013){Kaplan}, {Bhalerao}, {van Kerkwijk},
  {et~al.}}]{kaplan13}
{Kaplan}, D.~L., {Bhalerao}, V.~B., {van Kerkwijk}, M.~H., {et~al.} 2013, \apj,
  accepted

\bibitem[{{Kawka} {et~al.}(2010){Kawka}, {Vennes}, \& {Vaccaro}}]{kawka10}
{Kawka}, A., {Vennes}, S., \& {Vaccaro}, T.~R. 2010, \aap, 516, L7

\bibitem[{{Kenyon} \& {Garcia}(1986)}]{kenyon86}
{Kenyon}, S.~J. \& {Garcia}, M.~R. 1986, \aj, 91, 125

\bibitem[{{Kilic} {et~al.}(2007{\natexlab{a}}){Kilic}, {Allende Prieto},
  {Brown}, \& {Koester}}]{kilic07}
{Kilic}, M., {Allende Prieto}, C., {Brown}, W.~R., \& {Koester}, D.
  2007{\natexlab{a}}, \apj, 660, 1451

\bibitem[{{Kilic} {et~al.}(2010{\natexlab{a}}){Kilic}, {Allende Prieto},
  {Brown}, {et~al.}}]{kilic10b}
{Kilic}, M., {Allende Prieto}, C., {Brown}, W.~R., {et~al.} 2010{\natexlab{a}},
  \apjl, 721, L158

\bibitem[{{Kilic} {et~al.}(2010{\natexlab{b}}){Kilic}, {Brown}, {Allende
  Prieto}, {Kenyon}, \& {Panei}}]{kilic10}
{Kilic}, M., {Brown}, W.~R., {Allende Prieto}, C., {Kenyon}, S.~J., \& {Panei},
  J.~A. 2010{\natexlab{b}}, \apj, 716, 122

\bibitem[{{Kilic} {et~al.}(2007{\natexlab{b}}){Kilic}, {Brown}, {Allende
  Prieto}, {Pinsonneault}, \& {Kenyon}}]{kilic07b}
{Kilic}, M., {Brown}, W.~R., {Allende Prieto}, C., {Pinsonneault}, M., \&
  {Kenyon}, S. 2007{\natexlab{b}}, \apj, 664, 1088

\bibitem[{{Kilic} {et~al.}(2009){Kilic}, {Brown}, {Allende Prieto}, {Swift},
  {Kenyon}, {Liebert}, \& {Ag{\"u}eros}}]{kilic09}
{Kilic}, M., {Brown}, W.~R., {Allende Prieto}, C., {Swift}, B., {Kenyon},
  S.~J., {Liebert}, J., \& {Ag{\"u}eros}, M.~A. 2009, \apjl, 695, L92

\bibitem[{{Kilic} {et~al.}(2011{\natexlab{a}}){Kilic}, {Brown}, {Allende
  Prieto}, {et~al.}}]{kilic11a}
{Kilic}, M., {Brown}, W.~R., {Allende Prieto}, C., {et~al.} 2011{\natexlab{a}},
  \apj, 727, 3

\bibitem[{{Kilic} {et~al.}(2012){Kilic}, {Brown}, {Allende Prieto},
  {et~al.}}]{kilic12a}
---. 2012, \apj, 751, 141

\bibitem[{{Kilic} {et~al.}(2011{\natexlab{b}}){Kilic}, {Brown}, {Hermes},
  {et~al.}}]{kilic11c}
{Kilic}, M., {Brown}, W.~R., {Hermes}, J.~J., {et~al.} 2011{\natexlab{b}},
  \mnras, 418, L157

\bibitem[{{Kilic} {et~al.}(2011{\natexlab{c}}){Kilic}, {Brown}, {Kenyon},
  {et~al.}}]{kilic11b}
{Kilic}, M., {Brown}, W.~R., {Kenyon}, S.~J., {et~al.} 2011{\natexlab{c}},
  \mnras, 413, L101

\bibitem[{{Koester}(2008)}]{koester08}
{Koester}, D. 2008, ArXiv:0812.0482

\bibitem[{{Kurtz} \& {Mink}(1998)}]{kurtz98}
{Kurtz}, M.~J. \& {Mink}, D.~J. 1998, \pasp, 110, 934

\bibitem[{{Kurucz}(1993)}]{kurucz93}
{Kurucz}, R.~L. 1993, {SYNTHE Spectrum Synthesis Programs and Line Data}
  (Kurucz CD-ROM; Cambridge, MA: Smithsonian Astrophysical Observatory)

\bibitem[{{Liebert} {et~al.}(2005){Liebert}, {Bergeron}, \&
  {Holberg}}]{liebert05}
{Liebert}, J., {Bergeron}, P., \& {Holberg}, J.~B. 2005, \apjs, 156, 47

\bibitem[{{Marsh} {et~al.}(1995){Marsh}, {Dhillon}, \& {Duck}}]{marsh95}
{Marsh}, T.~R., {Dhillon}, V.~S., \& {Duck}, S.~R. 1995, \mnras, 275, 828

\bibitem[{{Marsh} {et~al.}(2004){Marsh}, {Nelemans}, \& {Steeghs}}]{marsh04}
{Marsh}, T.~R., {Nelemans}, G., \& {Steeghs}, D. 2004, \mnras, 350, 113

\bibitem[{{Massey} {et~al.}(1988){Massey}, {Strobel}, {Barnes}, \&
  {Anderson}}]{massey88}
{Massey}, P., {Strobel}, K., {Barnes}, J.~V., \& {Anderson}, E. 1988, \apj,
  328, 315

\bibitem[{{Michaud} {et~al.}(2008){Michaud}, {Richer}, \&
  {Richard}}]{michaud08}
{Michaud}, G., {Richer}, J., \& {Richard}, O. 2008, \apj, 675, 1223

\bibitem[{{Mullally} {et~al.}(2009){Mullally}, {Badenes}, {Thompson}, \&
  {Lupton}}]{mullally09}
{Mullally}, F., {Badenes}, C., {Thompson}, S.~E., \& {Lupton}, R. 2009, \apjl,
  707, L51

\bibitem[{{Nomoto}(1982)}]{nomoto82}
{Nomoto}, K. 1982, \apj, 253, 798

\bibitem[{{Paczy{\'n}ski}(1971)}]{paczynski71}
{Paczy{\'n}ski}, B. 1971, Acta Astron., 21, 1

\bibitem[{{Panei} {et~al.}(2007){Panei}, {Althaus}, {Chen}, \& {Han}}]{panei07}
{Panei}, J.~A., {Althaus}, L.~G., {Chen}, X., \& {Han}, Z. 2007, \mnras, 382,
  779

\bibitem[{{Paquette} {et~al.}(1986){Paquette}, {Pelletier}, {Fontaine}, \&
  {Michaud}}]{paquette86}
{Paquette}, C., {Pelletier}, C., {Fontaine}, G., \& {Michaud}, G. 1986, \apjs,
  61, 197

\bibitem[{{Pyrzas} {et~al.}(2012){Pyrzas}, {G{\"a}nsicke}, {Brady},
  {et~al.}}]{pyrzas12}
{Pyrzas}, S., {G{\"a}nsicke}, B.~T., {Brady}, S., {et~al.} 2012, \mnras, 419,
  817

\bibitem[{{Rebassa-Mansergas} {et~al.}(2012){Rebassa-Mansergas}, {Nebot
  G{\'o}mez-Mor{\'a}n}, {Schreiber}, {et~al.}}]{rebassa12}
{Rebassa-Mansergas}, A., {Nebot G{\'o}mez-Mor{\'a}n}, A., {Schreiber}, M.~R.,
  {et~al.} 2012, \mnras, 419, 806

\bibitem[{{Schmidt} {et~al.}(1989){Schmidt}, {Weymann}, \& {Foltz}}]{schmidt89}
{Schmidt}, G.~D., {Weymann}, R.~J., \& {Foltz}, C.~B. 1989, \pasp, 101, 713

\bibitem[{{Shen} \& {Bildsten}(2009)}]{shen09}
{Shen}, K.~J. \& {Bildsten}, L. 2009, \apj, 699, 1365

\bibitem[{{Silvotti} {et~al.}(2012){Silvotti}, {{\O}stensen}, {Bloemen},
  {et~al.}}]{silvotti12}
{Silvotti}, R., {{\O}stensen}, R.~H., {Bloemen}, S., {et~al.} 2012, \mnras,
  424, 1752

\bibitem[{{Sim} {et~al.}(2012){Sim}, {Fink}, {Kromer}, {et~al.}}]{sim12}
{Sim}, S.~A., {Fink}, M., {Kromer}, M., {et~al.} 2012, \mnras, 420, 3003

\bibitem[{{Steinfadt} {et~al.}(2010){Steinfadt}, {Kaplan}, {Shporer},
  {Bildsten}, \& {Howell}}]{steinfadt10}
{Steinfadt}, J.~D.~R., {Kaplan}, D.~L., {Shporer}, A., {Bildsten}, L., \&
  {Howell}, S.~B. 2010, \apjl, 716, L146

\bibitem[{{Tauris} {et~al.}(2012){Tauris}, {Langer}, \& {Kramer}}]{tauris12}
{Tauris}, T.~M., {Langer}, N., \& {Kramer}, M. 2012, \mnras, 425, 1601

\bibitem[{{Tremblay} \& {Bergeron}(2009)}]{tremblay09}
{Tremblay}, P.-E. \& {Bergeron}, P. 2009, \apj, 696, 1755

\bibitem[{{Tremblay} {et~al.}(2010){Tremblay}, {Bergeron}, {Kalirai}, \&
  {Gianninas}}]{tremblay10}
{Tremblay}, P.-E., {Bergeron}, P., {Kalirai}, J.~S., \& {Gianninas}, A. 2010,
  \apj, 712, 1345

\bibitem[{{Van Grootel} {et~al.}(2013){Van Grootel}, {Fontaine}, {Brassard}, \&
  {Dupret}}]{vangrootel13}
{Van Grootel}, V., {Fontaine}, G., {Brassard}, P., \& {Dupret}, M.-A. 2013,
  \apj, 762, 57

\bibitem[{{Vennes} {et~al.}(2009){Vennes}, {Kawka}, {Vaccaro}, \&
  {Silvestri}}]{vennes09}
{Vennes}, S., {Kawka}, A., {Vaccaro}, T.~R., \& {Silvestri}, N.~M. 2009, \aap,
  507, 1613

\bibitem[{{Vennes} {et~al.}(2011){Vennes}, {Thorstensen}, {Kawka},
  {et~al.}}]{vennes11}
{Vennes}, S., {Thorstensen}, J.~R., {Kawka}, A., {et~al.} 2011, \apjl, 737, L16

\bibitem[{{Woosley} \& {Weaver}(1994)}]{woosley94}
{Woosley}, S.~E. \& {Weaver}, T.~A. 1994, \apj, 423, 371

\end{thebibliography}

\appendix \section{DATA TABLE}

        Table \ref{tab:dat} presents our radial velocity measurements. The Table
columns include object name, heliocentric Julian date (based on UTC), heliocentric
radial velocity (uncorrected for the WD gravitational redshift), and velocity error.

\begin{deluxetable}{ccc}
\tabletypesize{\footnotesize}
\tablecolumns{3}
\tablewidth{0pt}
\tablecaption{Radial Velocity Measurements\label{tab:dat}}
\tablehead{
        \colhead{Object}& \colhead{HJD}  & \colhead{$v_{helio}$}\\
                        & (days$-$2450000) & (\kms )
}
        \startdata
J0056$-$0611 & 5864.778351 & $  308.7 \pm  9.7 $ \\
\nodata      & 5864.786174 & $  -64.5 \pm  9.5 $ \\
\nodata      & 5864.788281 & $ -162.4 \pm 10.7 $ \\
\nodata      & 5864.789600 & $ -249.1 \pm  8.8 $ \\
\nodata      & 5864.790804 & $ -258.9 \pm 10.8 $ \\
\nodata      & 5864.792007 & $ -304.9 \pm 10.3 $ \\
        \enddata

        \tablecomments{(This table is available in its entirety in
machine-readable and Virtual Observatory forms in the online journal. A portion is
shown here for guidance regarding its form and content.)}

\end{deluxetable}

\end{document}